\documentclass[pra,aps,amsmath,amssymb,superscriptaddress,twocolumn, longbibliography]{revtex4-1}
\usepackage{graphicx}
\usepackage{color,outlines}
\usepackage[english]{babel}
\usepackage{amsthm}
\usepackage[sc,osf]{mathpazo}
\usepackage{hyperref}
\usepackage{physics}
\usepackage{outlines}
\usepackage{tabularx}
\usepackage{mathtools}
\usepackage{lipsum}

\newcommand{\circlew}{
	\tikz[baseline=-0.5ex] \draw[fill=white] (0,0) circle (0.1);
}
\newcommand{\squarew}{
	\tikz[baseline=-0.5ex] \draw[fill=white] (-0.12,-0.12) rectangle (0.12,0.12);
}
\newcommand{\circleb}{
	\tikz[baseline=-0.5ex] \draw[fill=black] (0,0) circle (0.1);
}
\newcommand{\squareb}{
	\tikz[baseline=-0.5ex] \draw[fill=black] (-0.12,-0.12) rectangle (0.12,0.12);
}
\definecolor{green2}{RGB}{0,100,0}

\usepackage{tikz, pgfplots}
\usetikzlibrary{positioning}

\DeclarePairedDelimiter\floor{\lfloor}{\rfloor}

\begin{document}
	
	\title{Entanglement structure for finite system under dual-unitary dynamics}
	
	\author{Gaurav Rudra Malik}
	\email{gauravrudramalik.rs.phy22@itbhu.ac.in}
	\affiliation{Department of Physics, Indian Institute of Technology (Banaras Hindu University), Varanasi, India~221005} 
	
	\author{Rohit Kumar Shukla}
	\email{rohitkrshukla.rs.phy17@itbhu.ac.in}
	\affiliation{Department of Chemistry; Institute of Nanotechnology and Advanced Materials; Center for Quantum Entanglement Science and Technology, Bar-Ilan University, Ramat-Gan 5290002, Israel} 
	
	\author{Sudhanva Joshi}
	\email{sudhanvajoshi.rs.phy24@itbhu.ac.in}
	\affiliation{Department of Physics, Indian Institute of Technology (Banaras Hindu University), Varanasi, India~221005} 
	
	\author{S. Aravinda}
	\email{aravinda@iittp.ac.in}
	\affiliation{Department of Physics, Indian Institute of Technology Tirupati, Tirupati, India~517619} 
	
	\author{Sunil Kumar Mishra}
	\email{sunilkm.app@iitbhu.ac.in}
	\affiliation{Department of Physics, Indian Institute of Technology (Banaras Hindu University), Varanasi, India~221005}

	\begin{abstract}
		
		The dynamics of quantum many-body systems in the chaotic regime are of particular interest due to the associated phenomena of information scrambling and entanglement generation within the system. While these systems are typically intractable using traditional numerical methods, an effective framework can be implemented based on dual-unitary circuits which have emerged as a minimal model for maximally chaotic dynamics. In this work, we investigate how individual two-body operators influence the global dynamics of circuits composed of dual-unitaries. We study their effect on entanglement generation while examining it from both bipartite and multipartite perspectives. Here we also highlight the significant role of local unitaries in the dynamics when paired with operators from the dual-unitary class, showing that systems with identical entangling power can exhibit a range of differing entanglement growth rates. Furthermore, we present calculations establishing time-step-dependent lower bounds, which depend on both the initial state and the entangling power of the constituent operators. Finally, we find that time-evolving an initial state composed of pair products generates a state with nearly maximal multipartite entanglement content, approaching the bounds established by Absolutely Maximally Entangled (AME) states.
		
	\end{abstract}

	\maketitle
	\section{Introduction}
	\label{Introduction}
	
	Entanglement is a hallmark of quantum many-body systems and serves as a central resource in quantum information processing \cite{Entanglement_Horodecki, qinfo_and_comp, uni_quant_comp, ShorAlgo, Crypto} while also indicating important features within the system at hand, such as, the area law dependence which indicates the presence of a gapped spectrum \cite{Area_Law_Eisert}. It also has further implications for the numerical simulation of a given system via the applicability of tensor networks \cite{Orus_2014,dmrg_mps}.
	
	The dynamical nature of the system is also reflected in the generation of entanglement under time evolution. It is known that most quantum systems, in the absence of specific features, dynamically converge to the thermal state under unitary time evolution \cite{cmp_review_huse}. Here the initial specifications of the system are suppressed and the quantum observables may be accurately estimated using ensembles of random matrices \cite{Thermal_State}. A feature associated with the dynamics of a many-body system is information scrambling, which is observed in the presence of some dynamical complexity such as that brought on by nonintegrability \cite{Swingle_Scrambling, Prosen_nint, Prosen_nint2}. Apart from scrambling effects, nonintegrability also plays a prominent role in localization to thermalization transitions \cite{Abanin_int, Disordered_Ising}. There exist several metrics to quantify the scrambling of information within a system such as the bipartite mutual information (BMI), tripartite mutual information (TMI) \cite{BMI_TMI_Ref,shukla2023scrambling} and the out-of-time-order correlation (OTOC) \cite{OTOC_Lin, Rohit_PRB,shukla2022characteristic, Fortes_OTOC}. The OTOC shows exponential dependence with time for quantum chaotic systems associated with an infinite Hilbert space and is proportional to the quantum Lyapunov exponent \cite{Maldacena2016}.
	
	For nonintegrable systems having local interactions, unitary time evolution can be effectively simulated using random unitary quantum circuits and the associated growth of entanglement across equal bipartitions shows a linear dependence on time \cite{Nahum_RQC, Nahum_RQC2, Khemani_RQC, Sondhi_RQC}. The connection between dynamic complexity and entanglement growth is further substantiated with quantum chaotic time evolution leading to states with near-maximum entanglement, starting from a completely separable state \cite{Arul_Sir_JNB, Sarkar_Chaos}. Apart from state dependence, an analysis of the involved complexity can be made by studying the operators bringing about the said time evolution. The associated state-independent and local unitary invariant quantities include operator entanglement \cite{Zanardi_Operator_Entanglement} and entangling power \cite{Zanardi_Entangling_Power}. The latter plays a significant role in the dynamic properties of the system and can be effectively used to construct an ergodic hierarchy of quantum evolutions based on the rate of information scrambling \cite{Aravinda_Sir_PRR}, while the former has been shown to be intricately linked to the OTOC \cite{NamitAnand_OTOC_OE}.
	
	While the nonintegrable unitary evolution generates a rich set of physical features, their numerical simulations are challenging as the high rate of entanglement generation restricts the application of dimensionality reduction used in a tensor network approach. However, it is possible to obtain results for complex dynamics by means of dual-unitary operators, which serve as minimal models of generating quantum chaotic evolutions \cite{prosen_review_dual}. This already includes the spatiotemporal correlation functions \cite{Dual_Dynamics_Corr_Fun}, the OTOC \cite{Dual_Dynamics_OTOC} and entanglement membranes \cite{Suhail_Membrane}. The full state dynamics can also be obtained for the class of solvable states \cite{Dual_Dynamics_Solvable_MPS}, while the bipartite entanglement features can also be computed for a more generic class of states \cite{Dual_Dynamics_Foligno}. The computational power of the said circuits may be used for a quantum advantage \cite{Dual_Dynamics_Suzuki} and elements of ergodicity breaking are also observed along with scar states \cite{dual_ergodicity_breaking, Dual_Dynamics_Scars}.
	
	In our present work we consider the dynamics resulting from a finite width brickwall quantum circuit made up of dual-unitary operators acting on a generic initial state. Our motivation is to understand the manner in which an individual operator used in constructing the brickwall circuit influences the overall dynamics, particularly in reference to the entanglement structure generated within the system. In this regard, the entangling power of the single operator plays an important role with respect to both the bipartite and multipartite entanglement measures. We also consider the role of local operators, in terms of random single qubit unitaries, for the bipartite case which is known to affect the entanglement generating capabilities of a bipartite gate \cite{Bhargavi_Jonnadula2}. 
	
	For identifying the role of local dynamics we consider an ensemble of circuits where each member is composed of a repeating pattern of a single operator. Between members of the ensemble we use operators that are different, albeit local unitary equivalent. In analysing the results obtained, we establish the role of a single two-body operator in the overall system dynamics for a more generic case, left ambiguous by the entangling power. Our results also highlight the initial buildup of entanglement within a circuit of finite spatial width and temporal depth, and agree well with the said dynamics for an infinite system in both space and time \cite{Dual_Dynamics_Foligno}. Further, while there have been several studies based on the entanglement associated with bipartitions, not a lot of literature exists when considering multipartite entanglement \cite{Rajrishi_Pal_MultiEP, Arul_Sir_MultiEP}. In our results we again find a strong dependence of the entangling power of the individual gates and multipartite features of entanglement with respect to the measure of time-evolved state and also of the entire circuit. Overall we find that the entanglement generation is favored by chaotic dynamics, for both the bipartite and multipartite cases, similar to that in the semiclassical regime \cite{SohiniGhose_Entanglement_Chaos}.
	
	This manuscript is organized as follows: In Section \ref{setup} we discuss dual-unitary quantum circuits and the measures of mixing rate, its dependence on entangling power and the relationship with ergodic hierarchy. Section \ref{bipartite_entanglement_velocity} presents our results on bipartite entanglement growth, with a focus on the roles of mixing rate and entangling power. In Section \ref{multipartite_entanglement} we examine multipartite entanglement in the time-evolved state and characterize the circuit properties responsible for the observed entanglement structure. Comparative results for spin models are discussed in Section \ref{spin_models}, and we conclude with a summary of our findings in Section \ref{summary}.
	
	\section{Setup}
	\label{setup}
	\subsection{Dual Unitary Operators and Initial State}
	\label{Section_Operators_and_def}
	
	The quantum circuit central to our discussion is composed of unitary operators \( \hat{U} \) acting on two qudits of local dimension \( q \), specifically from the class of dual-unitary operators ($\mathcal{DU}(q)$). A unitary operator $\hat{U} \in \mathcal{DU}(q)$  if in addition to satisfying the standard unitary condition (\( \hat{U}\hat{U}^{\dagger} = \hat{U}^{\dagger}\hat{U} = I \)), its associated dual operator \( \tilde{\hat{U}} \) is also unitary. The two-body operator can be written as a rank-4 tensor as $U_{ijkl}$, where each index can take $q$ values $\{ 0,1,2,...q-1 \}$ . The particular element $U_{ijkl}$ is given by the expression:
	\begin{equation}
		U_{ijkl} = \bra{ij}\hat{U}\ket{kl}
		\label{dual_tensor_form}
	\end{equation}
	The dual operator \( \tilde{\hat{U}} \) is obtained by a specific rearrangement of the indices of the original operator tensor, known as the $R_1/R_2$ transformation. This operation, along with that partial transposition are defined via exchange transformations given by:
	
	\begin{itemize}
		\item $R_1$ realignment: $\bra{ij}\hat{U}\ket{kl} = \bra{lj}\hat{U}^{R_1}\ket{ki}$
		\item $R_2$ realignment: $\bra{ij}\hat{U}\ket{kl} = \bra{ik}\hat{U}^{R_2}\ket{jl}$
		\item $T_1$ partial transpose: $\bra{ij}\hat{U}\ket{kl} = \bra{kj}\hat{U}^{T_1}\ket{il}$
		\item $T_2$ partial transpose: $\bra{ij}\hat{U}\ket{kl} = \bra{il}\hat{U}^{T_2}\ket{jk}$
	\end{itemize}
	These definitions are useful from a notational perspective when considering dual-unitaries. We may now formally define $\mathcal{DU}(q)$ as:
	\begin{equation}
		\mathcal{DU}(q) = \{ \hat{U} \in \mathcal{L}(\mathcal{H}_q \otimes \mathcal{H}_q) \ | \ \hat{U}^{R} (\hat{U}^{R})^{\dagger} = \mathbb{I} = (\hat{U}^{R})^{\dagger} \hat{U}^{R}  \}
	\end{equation}
	Here, $R$ maybe taken as either $R_1$ or $R_2$, and $\mathcal{L}(V)$ represents all the unitary operators acting on the vector space $V$. Finally, $\mathcal{H}_q$ represents the Hilbert space of local dimension $q$. The dual operator $\tilde{\hat{U}}$ is given as $\hat{U}^{R}$. Moreover, there exists another class of operators called as 2-unitaries, $\mathcal{U}_2(q)$ given by:
	\begin{equation}
		\mathcal{U}_2(q) = \{ \hat{U} \in \mathcal{DU}(q) \ | \ \hat{U}^{T} (\hat{U}^{T})^{\dagger} = \mathbb{I} = (\hat{U}^{T})^{\dagger} \hat{U}^{T}  \}
	\end{equation}
	here, $T$ maybe taken as either $T_1$ or $T_2$. Clearly, $\mathcal{U}_2(q) \subset \mathcal{DU}(q) \subset \mathcal{L}(\mathcal{H}_q \otimes \mathcal{H}_q)$. For an operator $\hat{U} \in \mathcal{L}(\mathcal{H}_q \otimes \mathcal{H}_q)$, the implementation complexity depends on its non-locality which is estimated following the Schmidt decomposition of the operator into operators on constituent vector spaces:
	\begin{equation}
		\hat{U} = \sum_{j = 0}^{q^2 - 1} \sqrt{\gamma_j}\hat{X}_j \otimes \hat{Y}_j \,\,\,\,\,.
	\end{equation}
	For a separable operator the above decomposition has a single tensor product term, which maybe implemented in terms of the involved operators. For an entangled operator, a series of local operations are required, thereby signaling the greater implementation complexity as more number of gates are required to implement the said operation. The operators $\hat{X}_k$ and $\hat{Y}_k$ $\in \mathcal{L}(\mathcal{H}_q)$ form the orthonormal basis under the Hilbert-Schmidt (HS) product $\bra{\hat{A}_i}\hat{A}_j\rangle = \tr(\hat{A}_i^\dagger \hat{A}_j) = \delta_{ij}$. This follows as the space $\mathcal{L}(\mathcal{H}_q)$ along with the HS product itself forms a Hilbert space $\mathcal{H}^q_{HS}$, which is isomorphic to $\mathcal{H}_q^{\otimes 2}$. Following this, the entanglement within operator $\hat{U}$ is given as\cite{Operator_Schmidt_Nielsen}:
	\begin{equation}
		E(\hat{U}) = 1 - \frac{1}{q^4}\sum_{j = 0}^{q^2 - 1} \gamma_j^2 \,\,\,\,\,.
	\end{equation}
	In order to address the non-local and entangling nature of the classes $\mathcal{DU}(q)$ and $\mathcal{U}_2(q)$ from a state perspective, we may associate the unitary bipartite operation $\hat{U}$ to the four party state $\ket{\psi_{PQRS}}$ via the Choi-Jamiolkowski isomorphism between $\mathcal{U}(\mathcal{H})$ and $\mathcal{H} \otimes \mathcal{H}$, defined as \cite{Bengtsson_Zyczkowski_2006}:
	\begin{equation}
		\ket{\psi_{PQRS}} = (\hat{U}_{PQ} \otimes \mathbb{I}_{RS})\ket{\phi_{PR}^{+}} \otimes \ket{\phi_{QS}^{+}}\,\,\,\,\,.
		\label{four_party}
	\end{equation}
	This state is maximally entangled across the bipartition $PQ|RS$ owing to the unitary nature of the matrix $\hat{U}$. For a unitary matrix $\hat{U} \in \mathcal{DU}(q)$, maximal entanglement additionally exists for the bipartition $PR|QS$ and also $PS|RQ$ if $\hat{U} \in \mathcal{U}_2(q)$ \cite{Bhargavi_Jonnadula1} making $\ket{\psi_{PQRS}}$ an absolutely maximally entangled (AME) state \cite{AME_States}. While the AME states do not exist for arbitrary number of parties and local dimensions \cite{Euler_officers}, as for a four-party qubit state \cite{AME_Max_Ent}, examples do exist for local dimension $q = 3$ with the same number of qutrits. This implies that $\mathcal{U}_2(2) = \emptyset$ while $\mathcal{U}_2(3) \ne \emptyset$ following the correspondence established by Eq. \ref{four_party}. The maximum entanglement across the bipartition $PR|QS$ for $\ket{\psi_{PQRS}}$ also implies maximum operator entanglement $\forall \hat{U} \in \mathcal{DU}(q)$ as the reduced state $\rho_{PR} = \hat{U}^{R_1}(\hat{U}^{R_1})^{\dagger}/q^2$. With linearized entropy as the entanglement measure \cite{China_EP},
	\begin{equation}
		E(\hat{U}) = 1 - \tr({\rho_{PR}^2}) = 1 - \frac{1}{q^4}\tr\Big( \Big[\hat{U}^{R_1}(\hat{U}^{R_1})^{\dagger} \Big]^2 \Big).
	\end{equation}
	Thus, by definition of $\mathcal{DU}(q)$ and $\tr({\mathbb{I_q}}) = q^2$ we have:
	\begin{equation}
		E(\hat{U}) = max_U E(\hat{U}) \Big(= E(\hat{S}_q)\Big) = 1 - \frac{1}{q^2}, \forall \ \hat{U} \in \mathcal{DU}(q) 
		\label{max_eu}
	\end{equation}
	here $\hat{S}_q$ is the SWAP operator for local dimension $q$. In context of the $PS|RQ$ bipartition, the reduced state is given by $\rho_{PS} = \hat{U}^{T_2}(\hat{U}^{T_2})^{\dagger}/q^2$, which is same as $E(\hat{U}\hat{S})$. This can be understood by the action of $\hat{S}$ along with the operator $\hat{U}$ which leads to the state $\ket{\psi_{QPRS}}$ following Eq.(\ref{four_party}). Now, as $E(\hat{U})$ is given by $E(\ket{\psi_{PQRS}})_{PR|QS}$, it follows that $E(\hat{U}\hat{S}) = E(\ket{\psi_{QPRS}}_{QR|PS})$ and is the same as the expression obtained using the partial transpose operation $T$ \cite{Bhargavi_Jonnadula1, China_EP}. Hence we have $\forall \hat{U} \in \mathcal{U}_2(q)$, 
	\begin{equation}
		E(\hat{U}) = E(\hat{U}\hat{S}_q) = E(\hat{S}_q)\,\,\,\,\,,
		\label{condition_maxep}
	\end{equation} 
	which in turn maximizes the entangling power $e_P(\hat{U})$ i.e. the average bipartite entanglement generated when an ensemble of product initial states is acted upon by the operator $\hat{U}$, measured by a metric $\mathtt{E}$, and given as:
	\begin{equation}
		e_P(\hat{U}) \propto \ \mathbb{E}_{\psi_1,\psi_2} \Big[ \mathtt{E} \Big( \hat{U} \ (\ket{\psi_1} \otimes \ket{\psi_2}) \Big) \Big].
	\end{equation}
	This expression is an ensemble average which also has a closed form arising due to the unitary invariance of the Haar measure and group theoretic arguments \cite{Zanardi_Entangling_Power, Zanardi_Wang}. The same expression can also be had using an explicit linear algebra calculation \cite{Tutorial_Haar}, and given by:
	\begin{equation}
		e_P(\hat{U}) = C_q\Big( E(\hat{U}) + E(\hat{U}\hat{S}) - E(\hat{S}) \Big).
		\label{ep_eq}
	\end{equation}
	Here the factor $C_q = q^2/(q+1)^2$, implying that the maximum value that can be reached for $e_P(\hat{U})$ is $q^2E(\hat{S})/(q+1)^2$ when Eq. \ref{condition_maxep} is satisfied. This leads to $ e_p(\hat{U}) = (q-1)/(q+1)$ using Eq. (\ref{max_eu}) $\forall \hat{U} \in \mathcal{U}_2(q)$ \cite{Zanardi_Operator_Entanglement}. 
	
	In order to scale the values such that $0 \le e_P(\hat{U}) \le 1$, we divide the expression Eq. (\ref{ep_eq}) by the maximum $e_P(\hat{U})$ which also leads to the pre-factor taking the form $1/E(\hat{S})$. This defines the measure of $e_P(\hat{U})$ that we use within the manuscript. Thus, for the set $\mathcal{U}_2(q)$ the value of $e_P(\hat{U})$ is unity corresponding to the existence of the AME(4,$q$) state. Since the AME(4,2) state does not exist, a qubit operator cannot be 2-unitary, and therefore the entangling power cannot attain the value of 1. For qubits, the entangling power if therefore limited \cite{Kraus_Cirac} and can attain a maximum value of $2/9$ \cite{Balakrishnan_Sankaranarayan}, which following normalization is restricted to $2/3$. 
	
	Now using graphical notation to represent the circuit, we introduce the tensor form for a single local gate $\hat{U}$ and its hermitian conjugate $\hat{U}^{\dagger}$:
	\begin{equation}
		\begin{tikzpicture}[baseline=(current  bounding  box.center), scale=0.8]
			\draw[thick] (-4.25,0.5) -- (-3.25,-0.5);
			\draw[thick] (-4.25,-0.5) -- (-3.25,0.5);
			\draw[thick, fill=purple, rounded corners=2pt] (-4,0.25) rectangle (-3.5,-0.25);
			\node at (-4.75,0.05){$\hat{U}=$};
			\node at (-3.,-0.075){,};
		\end{tikzpicture}
		\qquad 
		\qquad
		\begin{tikzpicture}[baseline=(current  bounding  box.center), scale=0.8]
			\draw[thick] (-1.25,0.5) -- (-.25,-0.5);
			\draw[thick] (-1.25,-0.5) -- (-.25,0.5);
			\draw[thick, fill=teal, rounded corners=2pt] (-1,0.25) rectangle (-0.5,-0.25);
			\node at (-1.75,0.075){$\hat{U}^\dag=$};
			\node at (0,-0.075){.};
		\end{tikzpicture}
		\label{U_tensor}
	\end{equation}
	Here the four free legs indicate the free indices, that maybe contracted over to specify the unitary relations. We begin with primary unitary condition, given by $\hat{U}\hat{U}^{\dagger} = \hat{U}^{\dagger}\hat{U} = \mathbb{I}$, valid $\forall \ \hat{U} \in \mathcal{L}(\mathcal{H}_q \otimes \mathcal{H}_q)$:
	\begin{equation}
		\begin{tikzpicture}[baseline=(current  bounding  box.center), scale=0.5]
			
			\draw[thick] (-5.0,-1.0) -- (-3.5,0.5);
			\draw[thick] (-4.5, 1.5) -- (-3.0,3.0);
			\draw[thick] (3.0,-1.0) -- (4.5,0.5);
			\draw[thick] (3.5, 1.5) -- (5.0,3.0);
			
			\draw[thick] (-3.0, -1.0) -- (-4.5,0.5);
			\draw[thick] (-3.5, 1.5) -- (-5.0,3.0);
			\draw[thick] (5.0, -1.0) -- (3.5,0.5);
			\draw[thick] (3.0,3.0) -- (3.5,2.5);
			
			\node at (-2.0,1.05){$=$};
			\node at (2.0,1.05){$=$};
			
			\draw [thick] (-3.5,0.5) to[out=70, in=-70] (-3.5,1.5);
			\draw [thick] (-4.5,0.5) to[out=110, in=-110] (-4.5,1.5);
			\draw [thick] (3.5,0.5) to[out=110, in=-110] (3.5,1.5);
			\draw [thick] (4.5,0.5) to[out=70, in=-70] (4.5,1.5);
			
			\draw[thick] (-0.5, 2.5) -- (-0.5,-0.5);
			\draw[thick] (0.5, 2.5) -- (0.5,-0.5);
			\draw[thick] (-1.0, 3.0) -- (-0.5,2.5);
			\draw[thick] (-1.0, -1.0) -- (-0.5,-0.5);
			\draw[thick] (0.5, 2.5) -- (1.0,3.0);
			\draw[thick] (0.5, -0.5) -- (1.0,-1.0);

			\draw [thick,fill=purple,rounded corners=2.2pt] (-4.5,-0.5) rectangle (-3.5, 0.5);
			\draw [thick,fill=teal,rounded corners=2.2pt] (3.5,-0.5) rectangle (4.5, 0.5);
			\draw [thick,fill=teal,rounded corners=2.2pt] (-4.5,1.5) rectangle (-3.5, 2.5);
			\draw [thick,fill=purple,rounded corners=2.2pt] (3.5,1.5) rectangle (4.5, 2.5);

		\end{tikzpicture}.
	\end{equation}
	For every $\hat{U} \in \mathcal{DU}(q)$ the additional unitary condition, involving $\tilde{\hat{U}} = \hat{U}^R$, is satisfied via:
	\begin{equation}
		\begin{tikzpicture}[baseline=(current  bounding  box.center), scale=0.5]
			
			\draw[thick] (-5.5, 2.5) -- (-4.0,1.0);
			\draw[thick] (-3.0, 2.0) -- (-1.5,0.5);
			\draw[thick] (1.5, 2.5) -- (3.0,1.0);
			\draw[thick] (4.0, 2.0) -- (5.5,0.5);
			
			\draw[thick] (5.5, 2.5) -- (4.0,1.0);
			\draw[thick] (3.0, 2.0) -- (1.5,0.5);
			\draw[thick] (-1.5, 2.5) -- (-3.0,1.0);
			\draw[thick] (-4.0, 2.0) -- (-5.5,0.5);
			
			\draw [thick] (3.0,2.0) to[out=20, in=160] (4.0,2.0);
			\draw [thick] (3.0,1.0) to[out=-20, in=-160] (4.0,1.0);
			
			\draw [thick] (-3.0,2.0) to[out=160, in=20] (-4.0,2.0);
			\draw [thick] (-3.0,1.0) to[out=-160, in=-20] (-4.0,1.0);
			
			\node at (0,1.5){$=$};
			
			\draw [thick,fill=purple,rounded corners=2.2pt] (2,1) rectangle (3,2);
			\draw [thick,fill=teal,rounded corners=2.2pt] (4,1) rectangle (5,2);
			\draw [thick,fill=purple,rounded corners=2.2pt] (-2,1) rectangle (-3, 2);
			\draw [thick,fill=teal,rounded corners=2.2pt] (-4,1) rectangle (-5,2);
			
			\draw[thick] (-1.5, -1.5) -- (1.5, -1.5);
			\draw[thick] (-1.5, -2.5) -- (1.5, -2.5);
			
			\node at (-3.5,-2.0){$=$};
			
			\draw[thick] (-1.5, -1.5) -- (-2.0,-1.0);
			\draw[thick] (-1.5, -2.5) -- (-2.0, -3.0);
			\draw[thick] (1.5, -1.5) -- (2.0, -1.0);
			\draw[thick] (1.5, -2.5) -- (2.0, -3.0);

		\end{tikzpicture}.
	\end{equation}
	Finally, for 2-unitary operators, i.e. $\forall \hat{U} \in \mathcal{U}_2(q)$:
	\begin{equation}  
		\begin{tikzpicture}[baseline=(current  bounding box.center),scale = 0.5]
			\draw [thick,fill=teal,rounded corners=2.2pt] (1,1) rectangle (2,2); 
			\draw [thick] (1,2) -- (0.5,2.3);
			\draw [thick] (2,2) to[out=70, in=-70] (2,3);
			\draw [thick] (0.5,0.5) -- (1,1);
			\draw [thick] (1,3) -- (0.5,2.7);
			\draw [thick] (0.5,0.5) to[out=110, in=-110] (0.5,4.5);
			
			\draw [thick,fill=purple,rounded corners=2.2pt] (1,3) rectangle (2,4);
			\draw [thick] (1,4) -- (0.5,4.5);
			\draw [thick] (2,4) -- (2.5,4.5);
			
			\draw [thick] (1,1) -- (0.5,0.5);
			\draw [thick] (2,1) -- (2.5,0.5);
		\end{tikzpicture} = 
		\begin{tikzpicture}[baseline=(current  bounding box.center),scale = 0.5]
			\draw [thick] (1,2) -- (0.5,2.3);
			\draw [thick] (2,2) to[out=70, in=-70] (2,3);
			\draw [thick] (0.5,0.5) -- (1,1);
			\draw [thick] (1,3) -- (0.5,2.7);
			\draw [thick] (0.5,0.5) to[out=110, in=-110] (0.5,4.5);
			
			\draw [thick] (1,4) -- (0.5,4.5);
			\draw [thick] (2,4) -- (2.5,4.5);
			
			\draw [thick] (1,1) -- (0.5,0.5);
			\draw [thick] (2,1) -- (2.5,0.5);
			
			\draw [thick] (1,2) -- (2,2) ;
			\draw [thick] (1,1) -- (2,1) ;
			\draw [thick] (1,3) -- (2,3) ;
			\draw [thick] (1,4) -- (2,4) ;
		\end{tikzpicture};
		\;\;\;\
		\begin{tikzpicture}[baseline=(current  bounding box.center),scale = 0.5]
			\draw [thick,fill=teal,rounded corners=2.2pt] (1,1) rectangle (2,2); 
			\draw [thick] (0.5,0.5) -- (1,1);
			\draw [thick] (2.5,0.5) to[out=70, in=-70] (2.5,4.5);
			\draw [thick] (2,3) -- (2.3,2.7);
			\draw [thick] (2,2) -- (2.3,2.3);
			\draw [thick] (1,2) to[out=110, in=-110] (1,3);
			
			\draw [thick,fill=purple,rounded corners=2.2pt] (1,3) rectangle (2,4);
			\draw [thick] (1,4) -- (0.5,4.5);
			\draw [thick] (2,4) -- (2.5,4.5);
			
			\draw [thick] (1,1) -- (0.5,0.5);
			\draw [thick] (2,1) -- (2.5,0.5);
		\end{tikzpicture} = 
		\begin{tikzpicture}[baseline=(current  bounding box.center),scale = 0.5] 
			\draw [thick] (0.5,0.5) -- (1,1);
			\draw [thick] (2.5,0.5) to[out=70, in=-70] (2.5,4.5);
			\draw [thick] (2,3) -- (2.3,2.7);
			\draw [thick] (2,2) -- (2.3,2.3);
			\draw [thick] (1,2) to[out=110, in=-110] (1,3);
			
			\draw [thick] (1,4) -- (0.5,4.5);
			\draw [thick] (2,4) -- (2.5,4.5);
			
			\draw [thick] (1,1) -- (0.5,0.5);
			\draw [thick] (2,1) -- (2.5,0.5);
			
			\draw [thick] (1,2) -- (2,2) ;
			\draw [thick] (1,1) -- (2,1) ;
			\draw [thick] (1,3) -- (2,3) ;
			\draw [thick] (1,4) -- (2,4) ;
		\end{tikzpicture}.
		\label{tdual}
	\end{equation}
	These unitary relations enable the efficient calculation of quantities like the spatio-temporal correlation functions \cite{Dual_Dynamics_Corr_Fun} and the OTOC \cite{Dual_Dynamics_OTOC}. For our calculations, we take examples of dual-unitaries $\in \mathcal{DU}(2)$ using the Cartan decomposition of 2-qubit gates \cite{Kraus_Cirac}, and for the qutrit case we obtain elements $\in \mathcal{DU}(3)$ using the algorithmic approach described in \cite{Ensembles_Dual_Unitaries}(see Appendix \ref{appendix_qubit_dual} and \ref{appendix_qutrit_dual}). We use the algorithmic approach for qutrits as we can get operators with a range of entangling powers, which is useful for our analysis.
	
	We now describe the structure of our finite quantum circuit consisting of $L$ (= even) qudits where the associated time evolution takes place via a brickwall style arrangement of individual $\hat{U} \in \mathcal{DU}(q)$. Using the tensor notation introduced previously, a single time-step of the brickwall circuit as $\mathcal{U}$ is defined as:
	\begin{equation}
		\mathcal{U} =
		\Bigg[ \bigotimes\limits_{i \in \mathcal{Z}_{\text{even}}} \hat{U}^{i,i+1} \Bigg]
		\cdot
		\Bigg[ \bigotimes\limits_{j \in \mathcal{Z}_{\text{odd}}} \hat{U}^{j,j+1} \Bigg]
	\end{equation}
	Here $\hat{U}^{pq}$ represents the operator $\hat{U}$ acting on the $p th$ and $q th$ lattice site. In our numbering convention we assign integer values moving from left to right, starting with $0$ for the left-most index. Also, periodic boundary conditions are assumed, implying the $L$th index is the same as the $0$th index. Using the graphical notation specified in Eq. \ref{U_tensor} we therefore have $\mathcal{U}$ for $L = 12$:
	\begin{equation}
		\mathcal{U} =
		\begin{tikzpicture}[baseline=(current  bounding  box.center), scale=0.3]
			
			\foreach \i in {0,...,4}
			{
				\draw [thick] (-3.5+4*\i,-1) -- (-7.5+4*\i,3); 
			}
			
			\draw [thick] (-7.5,-1) -- (-9.5,1);
			\draw [thick] (14.5,1) -- (12.5,3);
			\draw [thick, dashed] (14.5,1) -- (-9.5,1);
			
			\foreach \i in {0,...,5} 
			{
				\draw [thick] (-9.5+4*\i,-1) -- (-5.5+4*\i,3);
				\draw [thick,fill=purple,rounded corners=2.2pt] (-9+4*\i,0.5) rectangle (-8+4*\i,-0.5);
				\draw [thick,fill=purple,rounded corners=2.2pt] (-7+4*\i,2.5) rectangle (-6+4*\i,1.5);
			}
			\foreach \i in {0,...,11} 
			{
				\node at (-9.5+2*\i,-1.6){\i};
			}
			\node at (14.5,-1.6){12};
		\end{tikzpicture}.
		\label{operator_tensor}
	\end{equation}
	For a total of $t$ time steps, the total time evolution operator is thereby given as: $\mathbb{U}(t) = \mathcal{U}^t$, which comprises $2t$ layers of gates with our numbering convention. This also implies that every time step gives an even integer value along the $t$ axis. The remaining parameter is that of the initial wavefunction, and for this purpose we consider the generic pair product state, which is represented as the following tensor product:
	\begin{equation}
		| \psi \rangle_{initial} = \frac{1}{q^{N/2}} \Big( \sum_{i,j = 0} ^ {q-1} m_{ij}|ij\rangle \Big)^{\otimes N}\,\,\,\,\,,
	\label{initial_state_eq}
	\end{equation}
	where $N = L/2$ specifies the number of pair products. The coefficients $m_{ij}$ of a given pair product itself form a rank-2 tensor, along with its conjugate, with the constraint $\tr(m m^{\dagger}) = d$ applied for normalisation. The generic matrix $m$ leads up to a class of states which do not admit any specific feature, and in general is taken to be a matrix with random entries in this manuscript.
	\begin{equation}
		\centering
		\begin{tikzpicture}[baseline=(current  bounding  box.center), scale=0.8]
			\draw[thick] (-0.5,0.5) -- (0,0);
			\draw[thick] (0,0) -- (0.5,0.5);
			\node at (-0.5,1.0){$i$};
			\node at (0.5,1.0){$j$};
			\draw[thick, fill = purple] (0,0) circle (0.2);
			\node at (-1.25,0.05){$m \longrightarrow$};
			
			\draw[thick] (3.5,0.0) -- (4,0.5);
			\draw[thick] (4,0.5) -- (4.5,0.0);
			\node at (3.5,-0.5){$i$};
			\node at (4.5,-0.5){$j$};
			\draw[thick, fill = teal] (4,0.5) circle (0.2);
			\node at (2.5,0.05){$m^{\dagger} \longrightarrow$};
			
		\end{tikzpicture}.
	\end{equation}
	The initial state for $L = 12$ is therefore represented by the following tensor, with the periodic boundary constraint by which the index $L$ is equivalent to index $0$:
	\begin{equation}
		\ket{\psi_{init}} = \,\,\,
		\begin{tikzpicture}[baseline=(current  bounding  box.center), scale=0.5]
			\foreach \i in {0,...,5}
			{
				\draw[thick] (-0.5+2*\i,0.5) -- (0+2*\i,0);
				\draw[thick] (0+2*\i,0) -- (0.5+2*\i,0.5);
				\draw[thick, fill = purple] (0+2*\i,0) circle (0.2);
			}
			\foreach \i in {1,...,12} 
			{
				\node at (-1.5+\i,1.5){\i};
			}
		\end{tikzpicture}.
		\label{state_tensor}
	\end{equation}
	In this representation, $j$ takes the value $i+1$. Thus the time evolution within our system is represented by a repeating segment of the operator $\mathcal{U}$ with the lower most indices being plugged by the tensor $\ket{\psi_{init}}$ to leave no open index other than those on the other end of the tensor structure, representing the time evolved state. For this, the index labeled as $k$ within the state tensor in Eq. \ref{state_tensor} is contracted with the correspondingly labeled index $k$ within the operator tensor in Eq. \ref{operator_tensor}. For the final index, we use periodic boundary condition, and instead contract with the $0$ index. Once the time evolved state is obtained, we can calculate the quantities of interest.
	
\begin{figure}[h]
	\includegraphics[width=\linewidth]{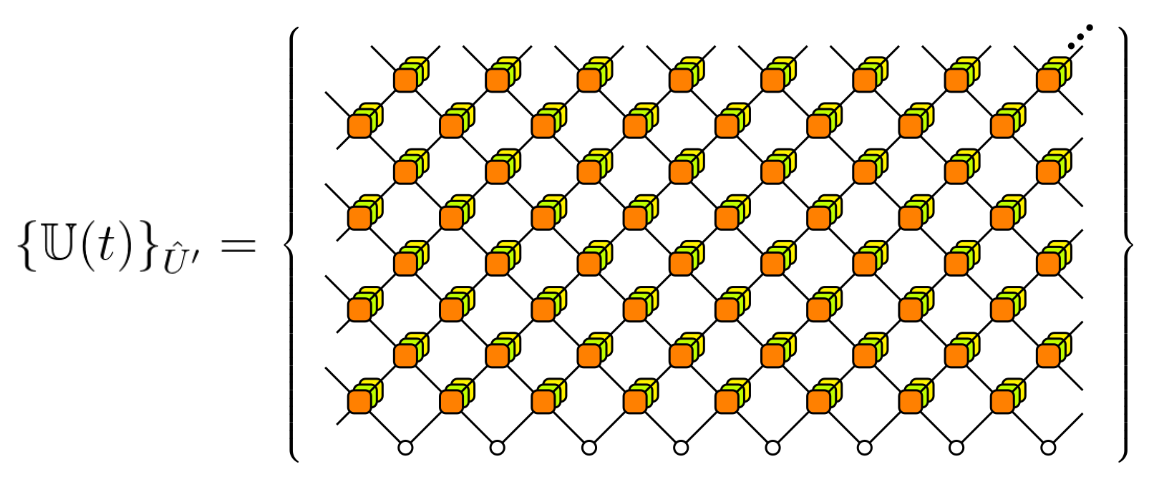}
	\caption{A schematic representation of the ensemble of circuits $\{ \mathbb{U}(t)\}_{\hat{U'}}$ with different $U'$ represented with blocks having different colours, all acting on similar initial states. The operators $U'$ represent different iterations of equation Eq. \ref{U_dash_figure}}
	\label{Fig_schematic}
	\end{figure}

	Our primary objective is to consider the effect of local operators $U$ on the global dynamics generated by the circuit $\mathbb{U}(t)$. The parameter of importance is therefore the entangling power $e_P(\hat{U})$. However using this metric it is not possible to account for local unitaries. To remedy this we take an ensemble of $\hat{U}'$ given by:
	\begin{equation}
		\hat{U}' = (\hat{u}_1 \otimes \hat{u}_2) \hat{U} (\hat{v}_1 \otimes \hat{v}_2),
		\label{U_dash_local_unitaries}
	\end{equation} 
	for a specified $\hat{U}$ having fixed entangling power, drawing the local unitary operators $u_i$ and $v_j$ from the Haar random ensemble for $i,j\in\{1,2\}$. The operator $\hat{U}'$ is represented graphically with a block having a different colour to that used in Eq. \ref{U_tensor}. For an particular example, this can be denoted in relation to $\hat{U}$ as:
	\begin{equation}
	\begin{tikzpicture}[baseline=(current  bounding  box.center), scale=0.5]
		\draw[thick] (0.5,0.5) -- (2.5,2.5);
		\draw[thick] (2.5,0.5) -- (0.5,2.5);
		\draw[thick, fill=orange, rounded corners=2pt] (1,1) rectangle (2,2);
		\node at (3.5,1.5) {$=$};
		\draw[thick] (4.5,0.5) -- (6.5,2.5);
		\draw[thick] (6.5,0.5) -- (4.5,2.5);
		\draw[thick, fill=purple, rounded corners=2pt] (5,1) rectangle (6,2);
		\draw[thick, fill = green] (4.5,0.5) circle (0.3);
		\draw[thick, fill = yellow] (6.5,2.5) circle (0.3);
		\draw[thick, fill = blue] (6.5,0.5) circle (0.3);
		\draw[thick, fill = black] (4.5,2.5) circle (0.3);
		\node at (5.5,0.5) {$u_1$};
		\node at (7.5,2.5) {$v_2$};
		\node at (7.5,0.5) {$u_2$};
		\node at (5.5,2.5) {$v_1$};
	
	\label{U_dash_figure}	
	\end{tikzpicture}
	\end{equation}
	here, $u_1,u_2$ and $v_1, v_2$ are represented by the dots of different colours, and are independently sampled. A block having with a separate colour is made up of a different, independently set of local unitaries while following the same graphical relation, for a specified $\hat{U}$. Note that $\hat{U}' \in \mathcal{DU}(q)$, when $\hat{U} \in \mathcal{DU}(q)$. Thereby replacing $U$ by different examples of $U'$ results in an ensemble of circuits $\{ \mathbb{U}(t)\}_{\hat{U}'}$ where we shall use a metric defined subsequently to highlight the role of local unitaries. Once $U'$ is specified, the circuit is Floquet in nature as the same sequence of gates is repeated for increasing number of time steps. A schematic diagram of our setup is given in Fig. \ref{Fig_schematic}. An ensemble of dual unitaries $U'$ generated by Haar-random local unitaries is created, leading to the collection $\{ \mathbb{U}(t)\}_{\hat{U}'}$.

	\subsection{Ergodic Hierarchy and Mixing Rate}
	\label{Section_Ergodic_hierarchy_mixing_rate}
	
	The correlation function between two space-time points in a $1+1D$ circuit composed of unitaries leads to a non-zero contribution iff the points lie within the light cone connecting them \cite{Chalker_minimal_models}. For the case where $\hat{U} \in \mathcal{DU}(q)$, the said correlations give a non-zero contribution only if the points lie exactly on the light cone and show decaying, oscillating or constant behavior with time. Further, for $\hat{U} \in \mathcal{U}_2(q)$ this correlation also vanishes, and non-zero behavior is found only in correlating a space-time point with itself. This range of dynamical behaviors for the correlation function for $\hat{U} \in \mathcal{DU}(q)$ is therefore indicative of an ergodic hierarchy with the class of $\mathcal{U}_2(q)$ operators at the apex. We shall begin with a review of known results which were originally discussed in references \cite{Aravinda_Sir_PRR, Dual_Dynamics_Corr_Fun}. These results are important to our present work and shall allow us to define measures which we shall make use of later in the manuscript. We include the previous known results to make our present work to be more self-contained.
	
	Using the notation from Ref. \cite{Dual_Dynamics_Corr_Fun}, the correlation function on the light cone for $\hat{U} \in \mathcal{DU}(q)$ is given as:
	\begin{equation}
		C^{\alpha \beta}_{\nu}(x = t,t) = \frac{1}{q}\tr[\mathcal{M}_{\nu}^{2t}(a^{\beta})a^{\alpha}].
		\label{corr_function}
	\end{equation}
	$x = t$ indicates the correlation function is evaluated at the light-cone $x = vt$ ($v = 1$ if $U \in \mathcal{DU}(q)$ \cite{Dual_Dynamics_OTOC}) with $\nu = \pm$ indicating the direction of light-cone. Here $a^{\alpha/\beta}$ denote the local observables. Thus the time dependence and the role of the operator both appear with the term $\mathcal{M}_{\pm}$, which can be defined for positive $\nu$ as follows:
	\begin{equation}
		\mathcal{M}_{+}(a) = \frac{1}{q} \tr_1 \Big[ \hat{U}^{\dagger}(a \otimes \mathbb{I}) \hat{U} \Big] = 
		\begin{tikzpicture}[baseline=(current  bounding  box.center), scale=0.5]
			
			\draw [thick] (0.5,-0.5) to[out=70, in=-70] (0.5,0.5);
			\draw [thick] (-0.5,-0.5) to[out=110, in=-110] (-0.5,0.5);
			\draw [thick] (-1.0,2.0) to[out=140, in=-140] (-1.0,-2.0);
			\draw[thick, fill = black](-0.6,0) circle (0.1cm);
			\node at (-1.05,0){$a$};
			
			\draw[thick] (-0.5,0.5) -- (1.0,2.0);
			\draw[thick] (0.5,0.5) -- (-1.0,2.0);
			\draw[thick] (0.5,-0.5) -- (-1.0,-2.0);
			\draw[thick] (-0.5,-0.5) -- (1.0,-2.0);
			
			\draw [thick,fill=purple,rounded corners=2.2pt] (-0.5,0.5) rectangle (0.5, 1.5);
			\draw [thick,fill=teal,rounded corners=2.2pt] (-0.5,-0.5) rectangle (0.5, -1.5);
			
		\end{tikzpicture}.
	\end{equation}
	Clearly, $\mathcal{M}_{+}$ is a unital channel that forms a linear map between an initial $\rho_i$ and final density matrix $\rho_f$, such that $\rho_f = \mathcal{M}_{+}(\rho_i)$. The said action can be modified following row vectorization as $M_{+}(\hat{U})$ acting on $\ket{\rho_i}$, which gives $\ket{\rho_f}$, such that $\ket{\rho_f} = M_{+}(\hat{U})\ket{\rho_i}$, where $\ket{\rho_{i/f}}$ are column vectors with $q^2$ entries. Following algebraic manipulation, we can also write the operators $M_{+}(\hat{U})$ in terms of matrix rearrangements \cite{Aravinda_Sir_PRR}
	\begin{equation}
		M_{+}(\hat{U}) = \frac{1}{q} \Big[ \hat{U}^{T_2}(\hat{U}^{T_2})^{\dagger} \Big]^{R_1}.
		\label{M_plus_eq}
	\end{equation}
	When $\nu$ is negative, $\mathcal{M}_{-}(a) = \frac{1}{q} \tr_2 \Big[ \hat{U}^{\dagger}(\mathbb{I} \otimes a) \hat{U} \Big]$, with $M_{-}(\hat{U}) = \frac{1}{q} \Big[ \hat{U}^{T_2}(\hat{U}^{T_2})^{\dagger} \Big]^{R_2}$ upon row vectorization.
	
	The nature of the correlation function in Eq. \ref{corr_function}, which is a reflection of the ergodic nature of the dynamics, depends strongly on the properties of eigenvalues of $M_{\pm}(\hat{U})$. This follows as a spectral decomposition of the matrix reveal the constant, decaying and oscillating modes. This in turn can be used to specify the ergodic hierarchy. Being unital in nature, one of the eigenvalues is certain to be unity, and forms a trivial case. We denote this eigenvalue as $\lambda_0$, and arrange the others for $M_{+}(\hat{U})$ (or $M_{-}(\hat{U})$ instead) in an increasing order based on their magnitude. This implies that $|\lambda_{q^2 - 1}| \le |\lambda_{q^2 - 2}| \le ... \le |\lambda_{2}| \le |\lambda_{1}| \le |\lambda_0| \ (= 1)$. The ergodic hierarchy is defined as follows:
	\begin{itemize}
		\item \textbf{Noninteracting}: all $(q^2 - 1)$ eigenvalues are $=1$. Constant correlations are obtained, an example is that of the $S_q$ gate.
		
		\item \textbf{Nonergodic}(Interacting and nonintegrable): $\exists k$, s.t. $\forall j > k \in \{1,2,...,q^2 - 1\}$, are non-trivial (i.e. $\lambda_j \neq 1$). Implies that $\lambda_l = 1 \ \forall \ l \le k$, which means that $k$ constant modes are present.
		
		\item \textbf{Ergodic but nonmixing}: $\lambda_k \neq 1 \forall \ k \in \{ 1,2,...q^2-1\}$. However, $\exists j \in \{ 1,2,...q^2-1\}$ s.t. $|\lambda_j| = 1$. This condition precludes mixing in the induced dynamics.
		
		\item \textbf{Ergodic and mixing}: All possible $(q^2 - 1)$ eigenvalues are within the unit circle, i.e. $|\lambda_k| < 1, \forall k \in \{ 1,2,...q^2-1\}$.
		
		\item \textbf{Bernoulli}: All $(q^2 - 1)$ eigenvalues have numerical value $= 0$.
		
	\end{itemize}
	Thus, assuming the ordering of eigenvalues according to their magnitude, the dynamical class in which a system belongs to can be judged by the value of $|\lambda_1|$, i.e. the largest non-trivial eigenvalue. Correspondingly we can also define an associated quantity called as the mixing rate $\mu_1 = -\log|\lambda_1|$ which directly corresponds to the ergodic nature of the system, being $0$ for the Non-Interacting class and going to $\infty$ for Bernoulli.
	\begin{figure*}
		\centering
		\includegraphics[scale = 0.95]{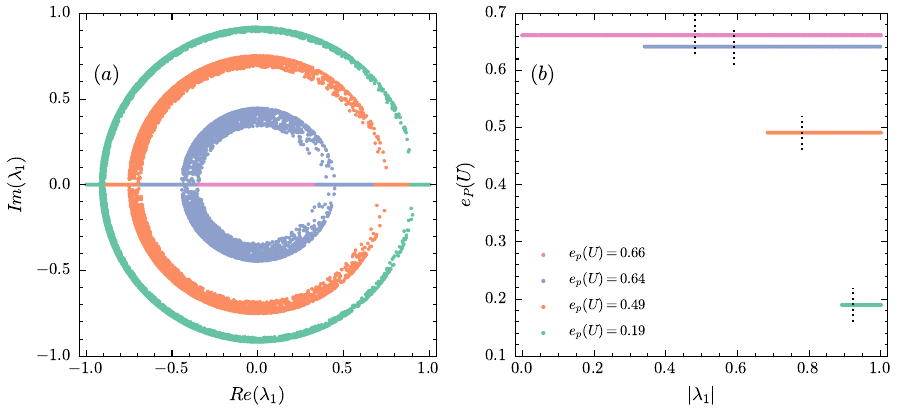}
		\centering
		\includegraphics[scale = 0.95]{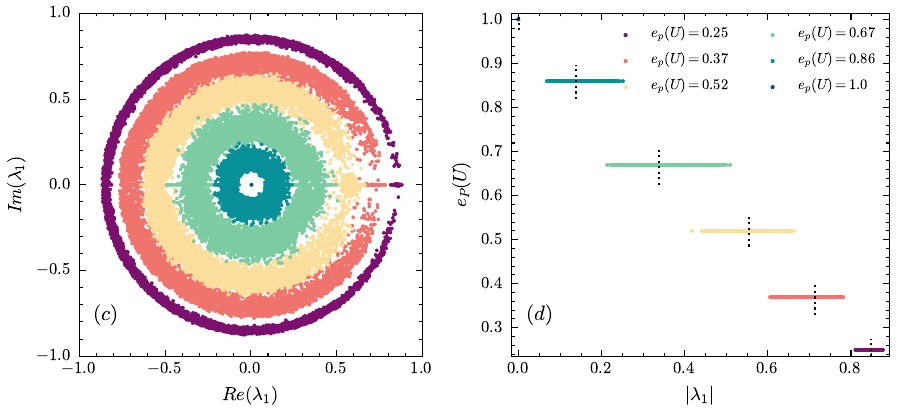}
		\caption{The variation of $\lambda_1$ with the entangling power $e_P(\hat{U})$, for an ensemble of 10000 $\hat{U}'$ operators created using Haar random local unitaries for qubits ($a$ and $b$) and qutrits ($c$ and $d$). In $(a)$ and $(c)$ we plot the real and imaginary parts of $\lambda_1$ for each member of the ensemble and in $(b)$ and $(d)$ we plot their magnitude, with the dotted line representing the average.}
		\label{fig_mixing_rate}
	\end{figure*}
	\begin{figure*}
		\centering
		\includegraphics[scale = 0.95]{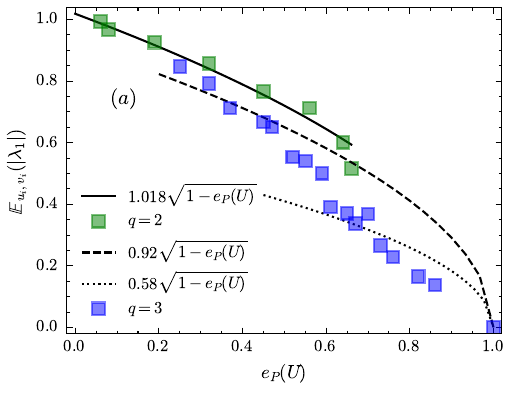}
		\centering
		\includegraphics[scale = 0.95]{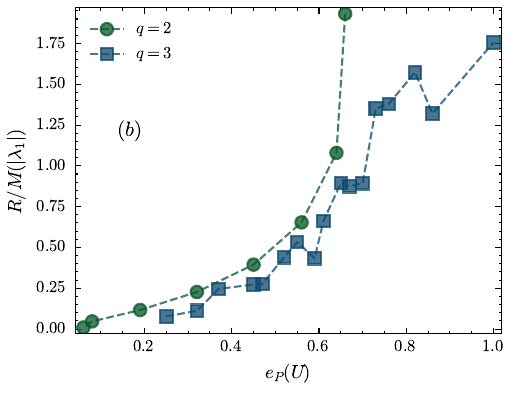}
		\caption{$(a)$ Variation for the average mixing rate $\mathbb{E}_{u_i,v_i} |\lambda_1|$ with entangling power $e_P(\hat{U})$ for the qubit and qutrit case, along with their respective fudge factors. $(b)$Variation of dispersion in $|\lambda_1|$ with $e_P(\hat{U})$}
		\label{fig_fudge_factor_and_dispersion}
	\end{figure*}
	
	Here it is important to note that the spectrum of $M_{\pm}(\hat{U})$ is not invariant upon the action of local unitaries. In fact for $\hat{U}^\prime = (\hat{u}_1 \otimes \hat{u}_2) \hat{U} (\hat{v}_1 \otimes \hat{v}_2)$, we have:
	\begin{equation}
		M_{+}(\hat{U}') = (\hat{v}_2^{\dagger} \otimes \hat{v}_2^{T})M_{+}(\hat{U})(\hat{u}_1^{\dagger} \otimes \hat{u}_1^{T}).
	\end{equation}
	Yet, the norm of $M_{\pm}(\hat{U})$ is constrained by the local unitary invariant of $e_P(\hat{U})$ as 
	\begin{equation}
		|| M_{+}(\hat{U}')||^2 = (q^2 - 1)(1 - e_P(\hat{U})) + 1,
		\label{norm_M}
	\end{equation}
	Thus for a given $\hat{U}$ with a fixed value of $e_P(\hat{U})$, an ensemble of $\hat{U}'$ operators may each have different values of the mixing rate. Further in context of Eq. \ref{norm_M} increasing the entangling power lowers the total sum of all eigenvalues, which tends to lower the value of $|\lambda_1|$. However, the distribution of $\lambda$ values is independent, leading to the ensemble having a wide range of mixing rates for a given value of $e_P(\hat{U})$.
	For the dual-circuit to be guaranteed as mixing, (i.e. $\abs{\lambda_1}<1$ irrespective of any random operators) the entangling power of the unitary brick $U$ must satisfy \cite{Aravinda_Sir_PRR}:
	\begin{equation}
		e_P(\hat{U}) > e_P^{*}(\hat{U}) = \frac{q^2 -2}{q^2-1}.
	\end{equation}
	In figures Fig. \ref{fig_mixing_rate}$(a), (c)$ we plot the real and imaginary part of $\lambda_1$ for an ensemble of local unitary equivalent operators and the range of magnitude obtained by the same for qubits and qutrits respectively in Fig. \ref{fig_mixing_rate}$(b), (d)$. We observe that for increasing values of $e_P(\hat{U})$, the ring formed by the real and imaginary parts of $\lambda_1$ becomes smaller, indicating diminished magnitudes of $\lambda_1$, which reduces to a point for the case when $e_P(\hat{U}) = 1$ in the qutrit case. This reflects the Bernoulli ergodic class corresponding to $|\lambda_i| = 0 \,\,\ \forall i$. Moreover, upon increasing the entangling power $e_P(\hat{U})$ the ensemble average of $\lambda_1$ get diminished, reflected in the lower reaches of $\lambda_1$ on increasing $e_P(\hat{U})$. Since $|\lambda_1|$ is the largest non trivial eigenvalue of the operator $M_{+}$, it maybe represented as the spectral radius, i.e. the largest magnitude eigenvalue, of an associated operator $\tilde{M}_+$, given as:
	\begin{equation}
		\tilde{M}_+ = M_+ - \ket{\phi}\bra{\phi}\,\,\,\,\,,
	\end{equation} 
	where $\ket{\phi}$ represents the eigenvector of $M$ having the trivial eigenvalue of $\lambda_0 = 1$. The eigen spectrum of the operator $\tilde{M}_+$ is given as $\{0, \lambda_{q^2-1}, \lambda_{q^2 - 2}, ..., \lambda_1 \}$ with the spectral radius for $\tilde{M}_+$ via Gelfand's formula being:
	\begin{equation}
		|\lambda_1| = lim_{k \rightarrow \infty} ||\tilde{M}_+^k||_{HS}^{1/k}\,\,\,\,\,,
		\label{Gelfand_eq}
	\end{equation} 
	where $||\tilde{M}_{+}^k||_{HS}$ represents the operator norm of $\tilde{M}_+$. Thus, the average mixing rate on the Haar random measure is given as:
	\begin{align}
		\mathbb{E}_{u_i,v_i} |\lambda_1| = \int_{Haar} du \ dv \lambda_1 \Big( (\hat{u}_1 \otimes \hat{u}_2)\hat{U}(\hat{v}_1 \otimes \hat{v}_2) \Big) \\
		= \int_{Haar} du \ dv \ \lambda_1 \Big( (\hat{u}_1*\hat{v}_1 \otimes \hat{u}_2*\hat{v}_2)\hat{U} \Big) \\
		= \int_{Haar} dw \lambda_1 \Big( (\hat{w} \otimes \hat{w}^{*})\hat{U} \Big),
	\end{align} 
	using Eq. \ref{Gelfand_eq}, we can equivalently write:
	\begin{equation}
		\mathbb{E}_{u_i,v_i} |\lambda_1| = lim_{k \rightarrow \infty} \int_{Haar} dw \lambda_1 \Big( ||(\hat{w} \otimes \hat{w}^{*})\hat{U}^k||_{HS}^{1/k} \Big).
	\end{equation}
	The Hilbert-Schmidt norm provides a bound for the magnitude of $|\lambda_1|$. Moreover the numerical value of $\overline{||\tilde{M}_+^k ||}_{HS}$ is approximated as \cite{Aravinda_Sir_PRR}:
	\begin{equation}
		\overline{||\tilde{M}_+^k ||}_{HS} \approx (1 - q^2)(1 - e_P(\hat{U}))^k\,\,\,\,\,.
	\end{equation}
	Using equation Eq. \ref{Gelfand_eq}, we can define an approximate bound for $\mathbb{E}_{u_i,v_i} |\lambda_1|$, with the help of a fudge factor $f$ to represent this relation.
	\begin{equation}
		\mathbb{E}_{u_i,v_i} |\lambda_1| \approx f\sqrt{1 - e_P(\hat{U}}) \,\,.
		\label{fudge_equation}
	\end{equation}
	The fudge factor clearly differentiates $e_P(\hat{U})$ from the average mixing rate $\mathbb{E}_{u_i,v_i}(\mu_1) = -\ln\Big( \mathbb{E}_{u_i,v_i} |\lambda_1| \Big) $ while establishing the approximate relation between the two. This fudge factor can be found by numerically fitting the  values of $\mathbb{E}_{u_i,v_i}$ with $\sqrt{1 - e_P(\hat{U})}$. For the case of qubits, defined via the Cartan form of dual unitaries there is a uniform fudge factor
	throughout the entire range of entangling powers. However, for the qutrit case where the operators are obtained via the $\mathcal{M}_R$ algorithm, different fudge factors are required for the higher and lower sections of $e_P(\hat{U})$ to minimize deviations (see \ref{fig_fudge_factor_and_dispersion} $(a)$). This highlights that while $|\lambda_1|$ certainly depends on $e_P(\hat{U})$, it is not a quantity that can be derived from it. Another feature that characterizes the relationship between mixing rate and entangling power is the relative dispersion of $|\lambda_1|$
	across different values of $e_P(\hat{U})$. We observe that lower entangling power $e_P(\hat{U})$ produces eigenvalues $|\lambda_1|$ with higher magnitudes and smaller variability, while higher $e_P(\hat{U})$
	leads to lower magnitudes with greater spread. To quantify this behavior, we define the relative dispersion measure:
	\begin{equation}
		\frac{R}{M}\Big( |\lambda_1| \Big) = \frac{\text{range}(|\lambda_1|)}{\overline{|\lambda_1|}}.
	\end{equation}
	This ratio increases as we move toward higher entangling power $e_P(\hat{U})$, primarily due to the decreasing mean value $\overline{|\lambda_1|}$ in the denominator. Since mixing rate is directly dependent on the value $|\lambda_1|$ we use the latter to describe the ergodic nature of the evolution. In the present case, the channels $\mathcal{M}_{\pm}$ come up following the structure of the brickwall circuit and simplifications brought on by dual-unitary properties of gates, which then define the ergodic nature of the circuit \cite{Aravinda_Sir_Channel, reference38}. Moreover, a general description can also be had for a channel denoting the evolution of a local operator under a bipartite unitary. By defining classical reduced channels from the Koopman operator, it can be shown that a quantum–classical correspondence emerges for the spectra of bipartite-unitary channels, where ergodicity governs the annular spectral bulk and mixing behavior \cite{correspondence}. The approach also yields classical analogues of dual-unitary circuits, with the associated dual-Koopman dynamics exhibit light-cone-restricted correlations and mixing behaviour governed by a similar contractive map which exists for circuits \cite{Arul_Sir_PRL}.
	
	Having reviewed some of the previously known results, we are now in a position to make use of the metrics for ergodicity that we have defined so far. In the following section we shall provide results relating the entanglement dynamics with the mixing rate within the system. This is an important observation given that the definition of channel $\mathcal{M}_{+}$ which leads to the matrix $M_{+}$ and its largest non-trivial eigenvalue $\lambda_1$ is in effect for an infinite temperature state at the thermodynamic limit. Yet we find that it has a discernible effect on the dynamics of a finite system with an arbitrary initial state. It therefore indicates mixing rate to be a fundamental feature related to the system dynamics.

	\section{Bipartite Entanglement Generation}
	\label{bipartite_entanglement_velocity}
	
	The entanglement growth within a system can be defined by the metric of entanglement velocity ($v_E$) \cite{Nahum_RQC2, Sarang_ent_val, Foligno2}. In the thermodynamic limit, $v_E$ defines the regime in which entanglement growth is linear in time. For a system under dual-unitary dynamics, this linear growth is maximal and implies that $v_E = 1$, provided that the entangling power of the gates constituing the circuit are finite \cite{Dual_Dynamics_Foligno} . Conversely, it has also been shown that the presence of maximal $v_E$ is indicative of dual-unitary dynamics \cite{converse}. Thus, it can be expected that for any non-trivial system defined by the initial state given by Eq. \ref{initial_state_eq}, the rate of entanglement generation eventually converges to maximal. When considering the finite system we can observe the entanglement growth prior to this convergence, and find that it highlights the ergodic properties of the circuit, eventually obscured in the intermediate time regime of linear growth. Thus, intial entanglement growth gives significant physical insights which we shall now discuss.
	
	Consider a subsystem denoted by $A$, comprising a series of $n$ contiguous qudits having indices: $i,i+1,...j$. The entanglement of this block with the remaining system $\bar{A}$ at a time step $t$ can be obtained by the entropy measures of the associated partial density matrix obtained after tracing out $\bar{A}$, with $\mathbb{U}(t) \in \{ \mathbb{U}(t)\}_{\hat{U}'}$:
	\begin{equation}
		\rho_A(t) = \tr_{\bar{A}} \Big( \mathbb{U}(t)\ket{\psi_{init}} \bra{\psi_{init}} \mathbb{U}^{\dagger}(t) \Big).
	\end{equation}
	The corresponding Renyi entropy of an order $\alpha \in \mathbb{R}$ can be calculated by:
	\begin{equation}
		S_A^{\alpha}(t) = \frac{1}{1 - \alpha} \ln \Big( \tr(\rho_A^{\alpha}(t)) \Big).
	\end{equation}
	To implement the expression $\rho_A^{\alpha}(t)$ by the graphical approach we proceed by drawing out the complete tensor diagram corresponding to $\ket{\psi(t)} = \mathbb{U}(t)\ket{\psi_{init}}$, and also for $\bra{\psi(t)} = \bra{\psi_{init}}\mathbb{U}(t)$ using conjugate elements. The partial trace can then be obtained by contracting indices corresponding to the subsystem $\bar{A}$, where using the unitary relations we can simplify the diagram. Taking powers of the resulting network, as required in calculating the Renyi entropy, the diagram factorises in terms of the initial state dependent $C_x$ matrix (see Appendix \ref{appendix_setting}) for a few initial time steps $t < t^{*}$($t^{*}$ is dependent on system size). The explicit tensor  representation of $C_x$ (involving a particular $\hat{U}'$) is given as \cite{Dual_Dynamics_Foligno}:
	\begin{equation}
		C_x = \frac{1}{q^{x/2}}
		\begin{tikzpicture}[baseline=(current  bounding  box.center), scale=0.3]
			
			\draw[thick] (-0.5,-0.5) -- (9,9);
			\draw[thick] (3.5,-0.5) -- (11,7);
			\draw[thick] (7.5,-0.5) -- (13,5);
			\draw[thick] (11.5,-0.5) -- (15,3);
			\draw[thick] (15.5,-0.5) -- (17,1);
			
			\draw[thick] (15.5,-0.5) -- (6,9);
			\draw[thick] (11.5,-0.5) -- (4,7);
			\draw[thick] (7.5,-0.5) -- (2,5);
			\draw[thick] (3.5,-0.5) -- (0,3);
			
			\draw[thick] (-0.5,-0.5) -- (-2.0,1.0);
			
			\foreach \i in {0,...,4} 
			{
				\draw [thick,fill=gray] (-0.5+4*\i,-0.5) circle (0.2);
			}
			
			\foreach \i in {0,...,3} 
			{
				\draw [thick,fill=gray,rounded corners=2.2pt] (1+4*\i,1) rectangle (2+4*\i,2);
			}
			\foreach \i in {0,...,2}
			{
				\draw [thick,fill=gray,rounded corners=2.2pt] (3+4*\i,3) rectangle (4+4*\i,4);
				
			}
			\foreach \i in {0,...,1}
			{
				\draw [thick,fill=gray,rounded corners=2.2pt] (5+4*\i,5) rectangle (6+4*\i,6);
			}
			\draw [thick,fill=gray,rounded corners=2.2pt] (7,7) rectangle (8,8);
			
		\end{tikzpicture}.
	\end{equation}
	The constant $1/q^{x/2}$ comes from the initial state where $1/\sqrt{q}$ is the normalisation for each of the involved $x$ number of pair products. The total open indices on the left and right hand sides of the tensor $C_x$ equals $2x$, implying that $C_x$ is matrix of order $q^x * q^x$. For simplicity we consider a half-half bipartition including $L/2$ qudits, such that $i = L/4$ and $j = L - i - 1$. For this construct, at a time $t$ the resulting $C$ operator is $C_{2t}$ ($C_{2t + 1}$) if $i$ is even (odd) factorising the $\tr(\rho^{\alpha}(t))$ expression as: \cite{Dual_Dynamics_Foligno}:
	\begin{equation}
		\tr(\rho^{\alpha}(t)) = \tr[(C_{\kappa_t}^{\dagger} C_{\kappa_t})^{\alpha}] \tr[(C_{\kappa_t}^{\dagger} C_{\kappa_t})^{\alpha}],
		\label{factor_rho}
	\end{equation}
	with $\kappa_t$ being the appropriate subscript for the operator $C$. For $L$ being an even number other than the multiple of $4$, the factors in Eq. \ref{factor_rho} would be different from each other. Also, for the factorisation to take place, the $C$ operators must not overlap, and this condition is given by $t < t^{*}$, where $t^{*}$ satisfies the constraint $2\kappa_{t^{*}} = N$. This can be observed as an operator $C_x$ has $x$ pair products, and at $t^{*}$ the operators are end-to-end, implying the constraint for $N$ number of total pair products. It indicates that $t^{*} = N/4$ or $N/4 - 1/2$ for $L$ being an even or odd multiple of $4$. Thus, we have:
	\begin{equation}
		S^{(\alpha)}_A(t) = \frac{2}{1 - \alpha}\ln \Big( \tr[(C_{\kappa_t}C_{\kappa_t}^{\dagger})^{\alpha}] \Big).
	\end{equation}
	
	Using the expression for $C_x$, it can be verified that for all cases $\tr (C_{x}C_{x}^{\dagger}) = 1$, with $\lambda_i$ being eigenvalues of $C_{x}C_{x}^{\dagger}$. With the constraint $\sum_i \lambda_i = 1$, we have:
	\begin{align}
		\frac{1}{(q^x)^{\alpha - 1}} \le \tr( (C_x C_x^{\dagger})^{\alpha}) \le 1 ,
		\\
		\implies 2x \ln q \ge \frac{2 \ln \tr( (C_x C_x^{\dagger})^{\alpha})}{1 - \alpha} \ge 0,
	\end{align}
	where the left (right) inequality is obtained for $\lambda_i = 1/q^x \, \forall i$ ($\lambda_i = 1$, for some $i$). For the maximal case, which corresponds to $m$ being unitary, the rate of entanglement growth coincides to the maximum value for all orders $\alpha$ \cite{Dual_Dynamics_Entanglement_Spreading}. The entanglement generated can then be quantified as a fraction of the maximum possible entanglement generated as a function of $x$, given by:
	\begin{equation}
		\mathcal{S}_A^{(\alpha)}(x) = \frac{2 \ln \tr( (C_x C_x^{\dagger})^{\alpha})/(1 - \alpha)}{2x\ln{q}} .
		\label{ent_vel_eq}
	\end{equation}
	The maximal increment of entanglement entropy between $t$ \& $t+1$ is $4\ln(q)$, irrespective of $\kappa_t$ \cite{Sarang_ent_val}. This gives us another metric to monitor entanglement growth as a fraction of the maximum, defined as:
	\begin{equation}
		\Delta S_A^{\alpha}(t) = \frac{S_A^{\alpha}(t) - S_A^{\alpha}(t-1)}{4\ln(q)}.
	\end{equation}
	Note that $\tr(C_{x}C_{x}^{\dagger}) = 1$ for all cases, but when $mm^{\dagger} = \mathbb{I}$, $C_{x}C_{x}^{\dagger}$ reduces to $\mathbb{I}_x/q^x$. This corresponds to the exactly solvable MPS \cite{Dual_Dynamics_Solvable_MPS} and the entanglement growth rate $\mathcal{S}^{(\alpha)}_{A}$ is found to be unity in all cases. The value of growth rate for an arbitrary $m$ does not reach the same for a finite case, and is dependent on circuit properties. 
	\begin{figure}
		\centering
		\includegraphics[scale = 0.95]{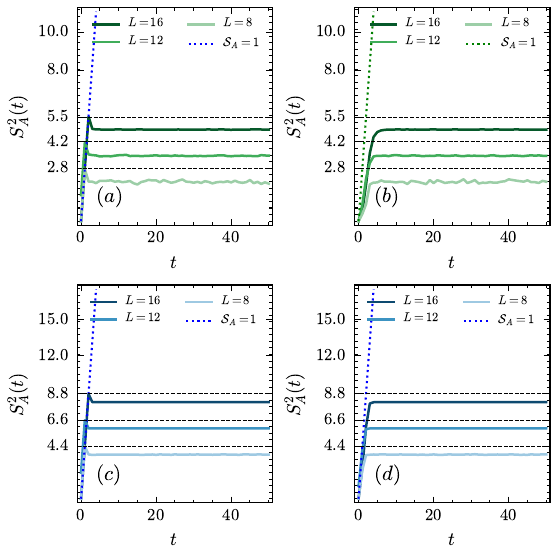}
		\caption{Build up of entanglement with time in the finite system for increasing sizes. $(a)$ \& $(b)$ denote the qubit case while $(c)$ \& $(d)$ are for qutrits. $(a)$ and $(c)$ denotes initial state with unitary $m$ and arbitrary entangling powers of $\hat{U}$, while $(b)$ and $(d)$ are for a random initial state, with $e_P(\hat{U}) = 0.61$ and $e_P(\hat{U}) = 0.86$ respectively. Note that the saturation of entanglement values is at the random matrix prediction, and slightly below the maximum possible entanglement for the respective system sizes denoted by the dotted lines.}
		\label{fig_finite_entanglement_growth}
	\end{figure}
	
	To calculate bipartite entanglement within our system for the $A|\bar{A}$ partition, we may alternatively generate the state $\mathbb{U}(t)\ket{\psi_{init}}$ evaluating $S_A^{\alpha}(t)$ numerically, in order to observe the entanglement saturation for the finite system. This approach gives us the same results in the regime $t \le t^{*}$ as those obtained via the $C$ operator, and while allowing us to probe further for larger $t$ where the factorization given by Eq. \ref{factor_rho} is no longer valid. Following this approach for $\alpha = 2$ we see in Fig. \ref{fig_finite_entanglement_growth} that the entanglement builds up rapidly, and saturates. This saturation value corresponds to the correspondingly sized random matrix ensemble. For the case of unitary $m$, the entanglement growth follows along the maximum rate of $\mathcal{S}_A^{2} = 1$, irrespective of the entangling power $e_P(\hat{U})$. Comparing this to the case where $m$ is randomly selected and non-unitary, the entanglement growth increases with increasing $e_P(\hat{U})$, edging closer to the case where $\mathcal{S}_A = 1$ (compare Fig \ref{fig_finite_entanglement_growth} $(b)$ and $(d)$). Taking examples of $N = 8, 12 \& 16$ we can see (in Fig \ref{fig_finite_entanglement_growth} $(a)$ and $(c)$) the different time-steps at which saturation occurs. Irrespective of $m$, the entanglement saturates at the same values for each $N$, with system size determining the saturation value as is expected from the respective random matrix distribution. It is just that for random $m$ is takes longer to reach there, depending on $e_P(\hat{U})$. Note that for the case of unitary $m$ the entanglement follows the maximal entanglement growth until $t = t^{*}$, before settling to the saturation value obtained by the corresponding random matrix calculations.
	
	Therefore, the role of single operator $\hat{U}'$ is most apparent when $t \le t^{*}$, for an initial state given by random $m$, defined for a finite system. To study that our approach is to create $C_{\kappa_t}$ operators with increasing values of $t$. This consideration implicitly assumes that the underlying finite system is large enough to satisfy the pre-saturation conditions as discussed earlier in this section, allowing us to probe entanglement generation prior to the regime of linear growth which is known to be maximal, even for arbitrarily selected $m$ matrix.
	
	Here it is important to note that the numerical values extracted for  the rate of entanglement growth for a finite time using Eq. \ref{ent_vel_eq}, are transient in nature. The metric we calculate is a normalised value of entanglement generated at time $t$, with the respect to the maximum possible value at the same time step and is given as $\mathcal{S}_A^{(\alpha)}(t) = S^{(\alpha)}_A(t)/S_A^{\text{max}}(t)$.   Evaluating this metric allows us to study the growth of entanglement in the initial few steps where the effects of dynamical ergodicity are most apparent, instead of the intermediate time regime where entanglement growth is linear in time and effects of initial states and nature of the individual operators are suppressed. This is true as even for arbitrary examples of initial states given by Eq. \ref{initial_state_eq}, where irrespective of $e_P(\hat{U})$, in the thermodynamic and long time limit, entanglement velocity reaches unity \cite{Dual_Dynamics_Foligno}. In other words, calculating the numerical values following Eq. \ref{ent_vel_eq} gives physical insight only in the finite case, as the convergence to maximum when the system extends in both space and time to infinity are well established. Taking $S_A^{\text{max}}(t)$ as $4t\ln(q)$, it can be seen that $\mathcal{S}$ has the same mathematical form as $v_E$, and is a finite system version of the quantity. The notational distinction is required as eventhough $\mathcal{S}_A^{(\alpha)}(t)$ has the same role as $v_E$ for a finite system, it does not represent the same physical situation as in the linear regime of entanglement growth. Note that normalised entanglement $\mathcal{S}$ is still a relevant quantity, having a proper structure and monotonic behaviour which also underlines the role of individual operators constructing the circuit. Making use of this metric we can monitor the buildup of entanglement in the finite system, and more importantly compare the extent of entanglement generated between different circuits having separate measures of ergodic indicators at a given time and a specific initial state. We proceed to do so in the following section, examining the values of $\mathcal{S}_A^{(\alpha)}(t)$ upon ensemble averaging and the average value of $|\lambda_1|$ for the ensemble which quantifies the average mixing rate, and find that on-average the increased ergodic mixing drives entanglement growth.
	
	\subsection{Dependence on Mixing Rate}
	\label{mixing_rate_dependence}
		\begin{figure*}	
		\centering
		\includegraphics[scale = 0.95]{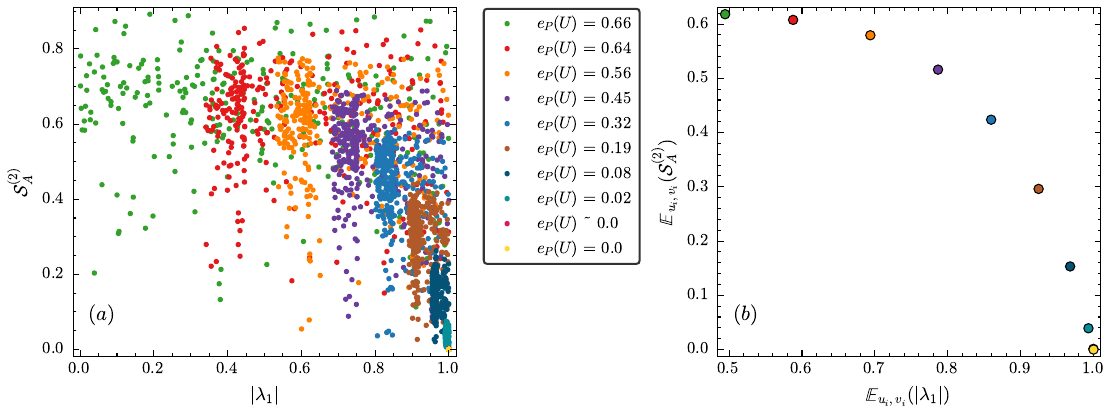}
		\includegraphics[scale = 0.95]{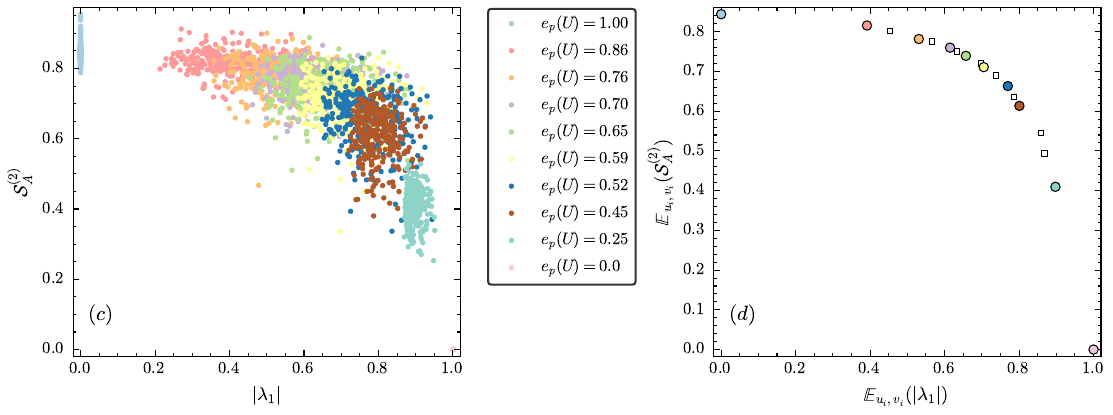}
		\caption{Values of normalised entanglement ($\mathcal{S}_A^{(2)}$) (for initial state $\ket{0}^{\otimes L}$) and mixing rates for an ensemble of 1000 $\hat{U}'$, at $t = 5$ and $\kappa_t = 2t$, for $(a)$ qubits and $(c)$ qutrits case along with the variation between $\mathbb{E}_{u_i,v_i}(\mathcal{S}_A^{(2)})$ and $\mathbb{E}_{u_i,v_i}(|\lambda_1|)$ for $(b)$ qubits and $(d)$ qutrits obtained for the same data upon ensemble averaging. $e_P(\hat{U})$ appears as a parameter characterising an ensemble. For a fixed value of $e_P(\hat{U})$, the associated data follows the same colour scheme for both individual members of the ensemble and corresponding averages (Compare parts $(a) \& (b)$ and $(c) \& (d))$. Note: Of the 1000 ensemble members, for each instance of $e_P(\hat{U})$, 300 are shown in the graphs on left hand side pane to avoid overcrowding, although averaging still involves all the 1000 ensemble members. Moreover in the qutrit case for some examples of $e_P(\hat{U})$ the individual ensemble members are not shown on the left pane, but the average is still calculated and shown as a unfilled square in part $(d)$.}
		\label{fig_mixing_rate_entanglement_velocity}	
	\end{figure*}
	
	In this section we present one of the main results of our manuscript, describing how the mixing rate influences entanglement dynamics as reflected in the measure of normalised entanglement ($\mathcal{S}_A^{(\alpha)}(t)$). We find that for an ensemble of operators $\hat{U}'$ constrained with a fixed entangling power there still exists a range of values for both $\mathcal{S}_A^{(\alpha)}(t)$ and $\Delta S_A^{\alpha}$ at a moderate time step $t$. This variation is found upon constructing the circuit operator $C_{\kappa_t}$ for all members of the $\hat{U}'$ ensemble and evaluating the normalised entanglement using equations Eq. \ref{factor_rho} and Eq. \ref{ent_vel_eq}, with the same initial state. We shall demonstrate that this variability is a result of local unitaries varying the mixing rate of the system while keeping the entangling power of individual gates unchanged. Finally, we find that it is the mixing rate which drives the growth of entanglement, revealing the impact of local unitaries in quantum circuit dynamics. 
	
	It is known that local unitaries have the tendency to alter the entanglement generating capabilities of a circuit, without affecting the $e_P(\hat{U})$ metric \cite{Bhargavi_Jonnadula1, Bhargavi_Jonnadula2}, but using a circuit of dual unitary operators allows us to explicitly state that the effect of local unitaries is manifested by their influence on the mixing rate. Note that $\lambda_1$ which determines the mixing rate is derived from the expression in Eq. \ref{corr_function}, which takes a solvable form only for the class of $\hat{U} \in \mathcal{DU}(q)$.
	
	For generating our results, we begin with an operator $\hat{U} \in \mathcal{DU}(q)$ for $q = 2$ and $3$, having a specific $e_P(\hat{U})$. We then create the ensemble $\{ \hat{U}' \}_{u,v}$ of operators, with each member leading to a distinct matrix $M_{+}$, and an associated value of $|\lambda_1|$. We have from Fig. \ref{fig_mixing_rate} $(b)$ and $(d)$ that increasing the value of $e_P(U)$ inversely affects the average over $|\lambda_1|$, however to observe its impact on $\mathcal{S}_A^{(\alpha)}(t)$ we construct the corresponding $C_{\kappa_t}$ operators for each member of the ensemble, and calculate $\mathcal{S}_A^{(\alpha = 2)}(t)$ given by the ratio $S^{(2)}_A(t)/4t\ln(q)$, taking $t = 5$. This process is repeated for a selection of $U$ giving a range of $e_P(\hat{U})$ values extending to unity for $\hat{U} \in \mathcal{DU}(3)$ for the case when $\hat{U} \in \mathcal{U}_2(3)$, and limited to $e_P(\hat{U}) = 0.66$ for $\hat{U} \in \mathcal{DU}(2)$. Having prepared the ensemble $\{ \hat{U}' \}_{u,v}$ and calculating the values of $\mathcal{S}_A^{(2)}$ along with $|\lambda_1|$, we plot them in a single graph to give Fig. \ref{fig_mixing_rate_entanglement_velocity} $(a)$ and $(c)$. We can clearly observe that for increasing values of $e_P(\hat{U})$, the points denoting ensemble members drift towards higher $\mathcal{S}^{(2)}_A$ and lower $|\lambda_1|$ values. 
	
	To quantify our finding in a more systematic manner we next plot the ensemble averages of the $\mathcal{S}_A^{(2)}$ and $|\lambda_1|$ that we obtained for the aforementioned ensemble $\{ \hat{U}' \}_{u,v}$ for each $e_P(\hat{U})$. This variation is presented in Figures Fig. \ref{fig_mixing_rate_entanglement_velocity} $(b)$ and $(d)$. While $e_P(\hat{U})$ does not have an explicit mention in the figure, each point in both the graphs represent an entire ensemble $\{ \hat{U}' \}_{u,v}$, having a fixed value of $e_P(\hat{U}') = e_P(\hat{U})$. In the overall picture, we find a monotonic increase in the normalised entanglement $\mathcal{S}_A^{(2)}$ upon the decrease in $|\lambda_1|$. A reduction of $\mathbb{E}(|\lambda_1|)$ from $0.9$ to $0.6$ is accompanied by a threefold increase in $\mathbb{E}(\mathcal{S}_A^{(2)})$ for qubit case. This increase is sustained to $\mathbb{E}_{u,v}|\lambda_1| = 0$ for the Bernoulli class in the case of qutrits, having $e_P(\hat{U}) = 1$. Here even a random $m$ gives a high average $\mathbb{E}(\mathcal{S}_A^{(2)})$ of $0.8$ for the qutrit Bernoulli case when $\lambda_1 = 0$ for all ensemble members. This indicates that while entanglement growth eventually achieves $v_E = 1$ for all cases, the circuit with higher ergodicity reaches that stage faster.
	
	At this point, it becomes important to stress that while our results till now consider the variation of two specific metrics viz. $\mathbb{E}(\mathcal{S}_A^{(2)})$ and $\mathbb{E}(\lambda_1)$ with respect to each other, entangling power still plays an implicit role in the overall picture, beyond affecting the mixing rate. Take the case when $e_P(\hat{U}) = 0$, there is no entanglement buildup, leading to a vanishing normalised entanglement $\mathcal{S}_A$. In the following section we elaborate on the role of entangling power, by presenting results for the explicit buildup of entanglement with time over the ensemble $\{ \hat{U}' \}_{u,v}$, taking the example of circuit operators $C_{2t}$ with increasing values of $t$.  
	
	\subsection{Role of Entangling Power}
	\label{entangling_power_dependence}
	\begin{figure*}
		\centering
		\includegraphics[scale = 1.0]{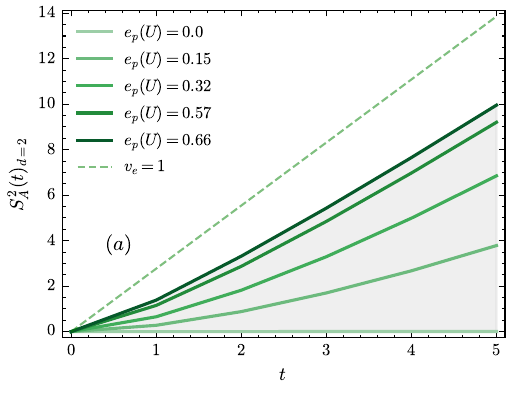}
		\includegraphics[scale = 1.0]{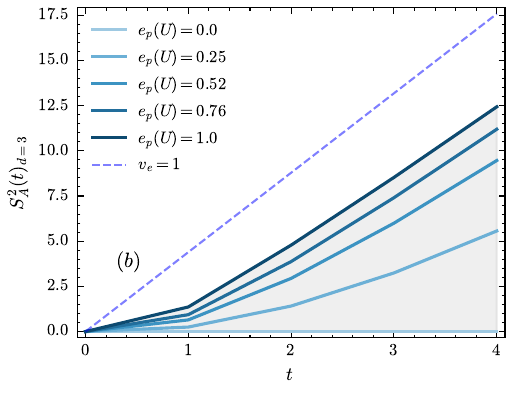}
		\caption{Growth of Renyi entropy, order $\alpha = 2$, for the $(a)$ qubit case and $(b)$ qutrit case with increasing time steps $t$ of the operator $C_{2t}$. The result is an ensemble average of 100 different initializations for the operator $\hat{U}'$ with the same randomly selected state. The shaded region gives an area wherein the average may lie for any arbitrary $e_P(\hat{U})$.}
		\label{fig_ent_velocity}
	\end{figure*}
	
	The correlation between entanglement velocity and mixing rate of the system implies that entanglement generation is an inherent feature of ergodic dynamics. Here entangling power $e_P(\hat{U})$ plays a crucial role, as it already influences the mixing rate as seen in Eq. \ref{fudge_equation}, and it is also known to lower bound the average entanglement growth in a brickwall circuit $\mathbb{U}$ \cite{Dual_Dynamics_Foligno}.  A similar lower bound appears for our Floquet setup as well, and upon calculation we shall see the role of entangling power. Although the calculations are exact for the model considered in \cite{Dual_Dynamics_Foligno}, our analysis is a close approximation which agree quite well for all time steps we could evaluate computationally (see Appendix \ref{appendix_entanglement_velocity}).   
	
	Taking $\hat{U}'$ as the building block of the circuit along with the local random unitaries we use the folded tensor notation to describe $\mathbb{E}_{u,v}\tr((C_{\kappa_t}C^{\dagger}_{\kappa_t})^2)$ as $\mathcal{P}_{\kappa_t}$ given here for $\kappa_t = 4$ as:
	\begin{equation}
		\mathcal{P}_{\kappa_t} = \,\,\,
		\begin{tikzpicture}[baseline=(current  bounding  box.center), scale=0.35]
			
			\foreach \i in {0,1,2,3}
			{
				
				\draw[thick] (-0.5+4*\i,0.5) -- (7+2*\i,8-2*\i);
				\draw[thick] (11.5-4*\i,0.5) -- (4-2*\i,8-2*\i);
			}
			
			\foreach \i in {0,1,2}
			{
				\draw[thick, fill=green, rounded corners=2pt] (1+4*\i, 2.0) rectangle (2+4*\i,3.0);
			} 
			
			\foreach \i in {0,1}
			{
				\draw[thick, fill=green, rounded corners=2pt] (3+4*\i, 4.0) rectangle (4+4*\i,5.0);
			} 
			\foreach \i in {0}
			{
				\draw[thick, fill=green, rounded corners=2pt] (5+4*\i, 6.0) rectangle (6+4*\i,7.0);
			}
			\foreach \i in {0,1,2,3}
			{
				\draw[thick, fill = green] (-0.5+4*\i,0.5) circle (0.2);
			}
			\foreach \i in {0,1,2,3}
			{	 
				\draw[thick, fill = white] (4-2*\i,8-2*\i) circle (0.3); 
			}
			\foreach \i in {0,1,2,3}
			{	 
				\draw[thick, fill = white] (7+2*\i,8-2*\i) rectangle (7.5+2*\i,8.5-2*\i); 
			}	
			
		\end{tikzpicture}.
	\end{equation}
	where we have the averaged gate:
	\begin{equation}
		\begin{tikzpicture}[baseline=(current  bounding  box.center), scale=0.5]
			\draw[thick] (0.5,0.5) -- (2.5,2.5);
			\draw[thick] (2.5,0.5) -- (0.5,2.5);
			\draw[thick, fill=green, rounded corners=2pt] (1,1) rectangle (2,2);
			\node at (6.5,2.5) {$= (U' \otimes_r {U'}^*)^{\otimes_r^2}$};
			\node at (9.0,0.5) {$= (P \otimes_r P) (U \otimes_r U^*)^{\otimes_r^2} (P \otimes_r P)$};
			
		\end{tikzpicture}.
	\end{equation}
	Here $\otimes_r$ denotes the tensor product over copies of $U'$ and ${U'}^*$ stacked over each other, obtained on folding the diagram $\tr((C_{\kappa_t}C^{\dagger}_{\kappa_t})^2)$ and $P$ denotes the projector operator arising out of averaging over haar random unitaries. The states $\ket{\circlew}$ and $\ket{\squarew}$ are given as:
	
	\begin{align}
		\ket{\circlew} = \frac{1}{q}\sum \delta_{ij}\delta_{kl}\ket{ijkl}, \\
		\ket{\squarew} = \frac{1}{q}\sum \delta_{il}\delta_{jk}\ket{ijkl}.
	\end{align}
	$\bar{S}_A^2(t)$ represents the average entanglement generated by an ensemble of operators $\hat{U}'$ and corresponding $C'_{\kappa_t}$ and is equal to  $\mathbb{E}_{u,v}\big[-2\ln(\tr[({C'}_{\kappa_t}{C'}_{\kappa_t}^{\dagger})^2])\big]$. By Jensen's inequality, we have for a convex function like $-2 \ln(x)$ that $\mathbb{E}(f(x)) \le f(\mathbb{E}(x))$, implying:
	\begin{equation}
		\bar{S}_A^2(t) \ge
		-2\ln\Big[\mathbb{E}_{u,v} \Big( \tr({C'}_{\kappa_t}{C'}_{\kappa_t}^{\dagger})^2 \Big)\Big].
	\end{equation}
	The term $\mathbb{E}_{u,v}\tr((C_{\kappa_t}C^{\dagger}_{\kappa_t})^2)$ on RHS equals $\mathcal{P}_{\kappa_t}$ which can be obtained analytically for small $\kappa_t$ (see Appendix \ref{appendix_entanglement_velocity} for details) and serves as the lower bound for the average entanglement generated within the system. 
	\begin{align*}
		\mathcal{P}_1 = \frac{c}{q}; \,\,\, c = \frac{\tr((mm^{\dagger})^2)}{q} \in [1,q] \,\,\,\,\,\,\,\,\,\,\,\,\,\,\,\,\,\,\,\,\,\,\,\,\,\,\,\  \\
		\mathcal{P}_2 = \frac{1}{q^2} + \frac{2\chi}{q^2}\sqrt{q^2 - 1}  + 
		\frac{2\chi^2}{q^2}(q^2(1-e_P) - 1) \\
		\mathcal{P}_3 \le  \frac{1}{q^3} + \frac{2\chi}{q^3}\sqrt{q^2 - 1}  + 
		\frac{2\chi^2}{q^3}(q^2(1-e_P) - 1) + \\ \Bigg( \chi^2\frac{q+c}{q+1}\sqrt{\frac{\eta^3(e_p)(q^2 - 1)}{q^2}} \Bigg)
	\end{align*} 
	here $\chi = (c-1)/\sqrt{q^2 - 1} \in \Big[0,\sqrt{\frac{q-1}{q+1}}\Big]$ is a quantity necessarily smaller than unity and $\eta(e_P)$ is rapidly decaying function on increasing $e_P$, and is unity at $e_P = 0$. 
	\begin{equation}
		\eta(e_P) = (1 - e_P)^2 + \frac{e_P^2}{q^2 - 1}
	\end{equation}
	$\mathcal{P}_3$ is the first non-trivial term, and simpifies into an inequality. Upon increasing $x > 3$ for $\mathcal{P}_x$ we get an expression involving higher powers of $\chi$. The term $\eta(e_P)$ has a negligible value for even moderately high $e_P$ and can be safely neglected. In the resulting expression we may extract the factor $1/q^x$ from $\mathcal{P}_x$ to give $\mathcal{P}_x = p_x/q^x$.
	\begin{equation}
		\implies -2\ln(\mathcal{P}_x) = 2x\ln(q) - \ln(p_x) \le \bar{S}_A^2(t).
	\end{equation}  
	The lower bound for average normalised entanglement (
	$\mathbb{E}(\mathcal{S}_A^{(2)}(x)$)) is obtained by dividing the above expression by the maximum value, given by $2x\ln(q)$:
	\begin{equation}
		\mathbb{E}(\mathcal{S}_A^{(2)}(x)) \ge 1 - \frac{\ln(p_x)}{2x\ln(q)}.
	\end{equation} 
	The term subtracted from unity has its numerator as logarithm of a summation with terms containing increasing power of $\chi$, while it is being divided by linearly increasing $2x\ln(q)$ factor. Hence, with each successive $x$, the entanglement velocity bound edges closer to unity. For the infinite case in terms of space and time, it is known that the value of $v_E(t)$ does approach unity $\forall U \in \mathcal{DU}(q)$ at $t \rightarrow \infty$, assuming sufficient entangling power associated with the gates $U$ forming the circuit \cite{Dual_Dynamics_Foligno}. This result holds for an arbitrary set of local unitaries at each space time point. Our setup for the finite case can be taken as an example of this, with the local unitaries $u$ and $v$ mounted upon $\hat{U}'$ in effect a two-site invariant collection of local-unitaries, thereby explaining our observations. A check on the expression is also provided, as taking $c = q$ (i.e. the case of unitary $m$) reduces each $p_x$ to $1$, thereby making the lower bound equal to unity $\forall x$. Further, taking $e_P = 0$ for a typical value of $c$ gives a negative lower bound, explaining the zero entanglement growth observed.
	\begin{figure*}
		\centering
		\includegraphics[scale = 0.95]{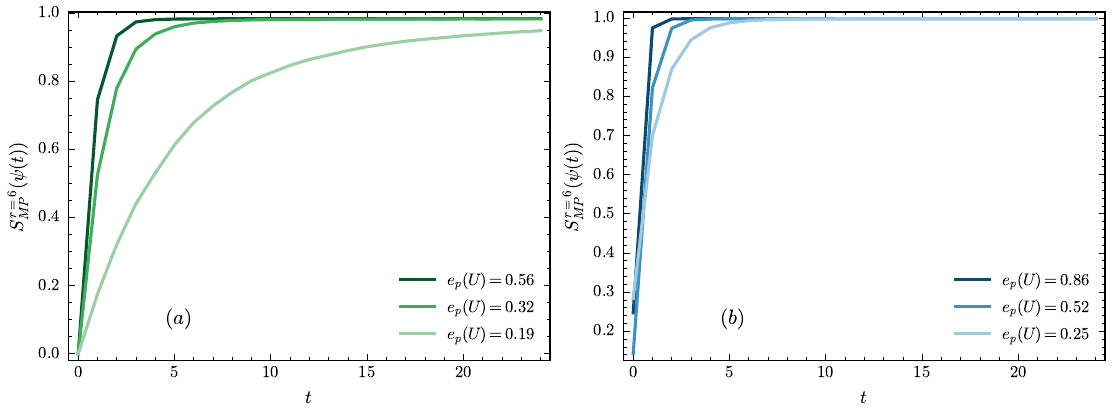}
		\caption{Variation of Scott Measure for $L = 12, r = 6$, for $(a)$ qubits, with initial state given by $\ket{0}^{\otimes L}$ i.e. $m = \text{diag}(1,0)$, $(b)$ qutrits, with initial state given by random $m$, having non-zero value of Scott Measure, with ensemble size for $\{\mathbb{U} \}_{\hat{U}'}$ = 100}
		\label{fig_multipartite_entanglement_qubit_qutrits}
	\end{figure*}
	
	While $e_P(\hat{U})$ sets a lower bound, the actual on-average entanglement growth significantly exceeds those. In the figures Fig. \ref{fig_ent_velocity} $(a)$ and $(b)$  we have calculated the entanglement values by selecting an operator $U$ having a specific value of $e_P$, and then generating an ensemble of operators $\hat{U}'$ by selecting matrices $u$ and $v$ from the haar random distribution (see Eq. \ref{U_dash_local_unitaries}). Following this we construct the $C_{2t}$ operator for increasing values of $t$ and the final result is the ensemble average over all operators. This plot helps us see the actual buildup of entanglement and its dependence on entangling power. In this setting the entanglement growth is found to be dependent on $e_P(\hat{U}')$ and scales positively with it upon increasing $t$. Our results align with the theoretical calculations of Ref. \cite{Dual_Dynamics_Foligno} as we see a gradual increase in entanglement increment $\Delta S_A^{\alpha}(t)$ (see Fig. \ref{fig_ent_velocity} $(a)$ and $(b)$) for each sucessive time step, as we take the operator $C_x$ with larger spatial and temporal dimensions. These successive increments presumably take $\mathbb{E} (\tilde{S}_A^{(2)})$ to unity for the infinite case, validating the results wherein $v_E = 1$.
	
	\section{Multipartite Entanglement}
	\label{multipartite_entanglement}
	
	We have observed for the bipartite case that increasing entangling power of the single gate, on-average, increases the entanglement generation of the system over an ensemble of $\hat{U}'$ operators which follows an average increase within the mixing rate. Now we consider the multipartite aspects of the generated entanglement structure and the role of single gate entangling power. We shall first discuss measures of multipartite entanglement and then we shall establish the aforementioned relation in reference to the time-evolved state. 
	
	In relation to quantum chaotic models, it is well known that with time there is rapid growth in bipartite entanglement across a given partition. Compared to that, the growth of multipartite entanglement is a more robust measure that generalises the entanglement across a bipartition to average entanglement across all possible bipartitions. Multipartite entanglement is more useful as a resource for implementing quantum algorithms \cite{multipartite_teleportation, multipartite_PRL} and also captures the spread of correlations near a quantum phase transition \cite{multipartite_scaling, first_order}. Moreover, examining growth of multipartite entanglement allows us to differentiate between classes ergodic circuits such as random unitary and random Clifford circuits \cite{GME_Random, mitsuhashi_prl}, where the saturation for a multipartite measure gives a stronger signature of thermalisation.
	
	\subsection{Measures for Multipartite Entanglement}
	\label{multipartite_measures}
	
	While there are several measures for bipartite entanglement, an exact apriori specification for multipartite entanglement is not well defined. Infact, most multipartite measures are obtained by averaging the bipartite entanglement over a given set of bipartitions. A simple example of that is the Mayer-Wallach $Q$ measure, which takes the following form \cite{Scott_Measure}:
	\begin{equation}
		Q(\psi) = 2\Big( 1 - \frac{1}{L}\sum_{k=1}^{L} \tr(\rho_k^2) \Big),
	\end{equation}  
	where $\rho_k$ represents the density matrix obtained after tracing out the $k$ th qudit. This measure calculates the average purity by iteratively tracing out each qubit and computing the mean across all single-qubit reductions. It may be generalised by the Scott Measure, taking the average over all possible $r$ sized bipartitions \cite{Scott_Measure}.
	\begin{equation}
		Q_r(\psi) = \frac{q^r}{q^r - 1} \Big( 1 - \frac{1}{^L C_r} \sum_{|S| = r} \tr(\rho_S^2)\Big).
	\end{equation}
	\begin{figure}
		\centering
		\includegraphics[scale = 0.95]{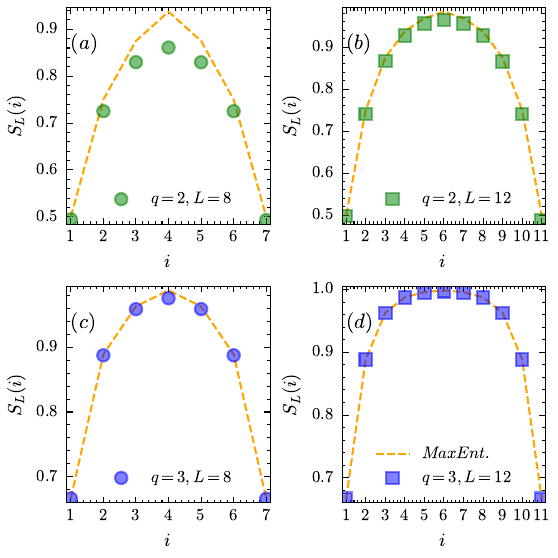}
		\caption{Comparison of entanglement via the linearised entropy at each bond of the time evolved state with its corresponding maximum value.}
		\label{fig_entanglement_and_bonds}
	\end{figure}
	
	In our analysis, we calculate the Scott measure at successive time steps using quantum circuits $\mathbb{U}(t)$ with $e_P(\hat{U})$ as a parameter, applied to initial states characterized by the matrix $m$. We denote this time-dependent measure as $S_{MP}^r(\psi(t))$. For our calculations, we set $r = L/2$, i.e. maximum partition size for the measure, and present results for both qubit and qutrit case. We examine two conditions: cases with zero initial value (Fig. \ref{fig_multipartite_entanglement_qubit_qutrits}(a)) and non-zero initial values (Fig. \ref{fig_multipartite_entanglement_qubit_qutrits}(b)), with the final result being an ensemble average over $\{\mathbb{U} \}_{\hat{U}'}$. The first case corresponds to using a diagonal matrix $m = \text{diag}(1,0)$, resulting in initial state $\ket{0}^{\otimes L}$ while the latter employs a matrix $m$ with random entries. Our results demonstrate that multipartite entanglement grows at a rate that scales directly with the entangling power of the constituent operators, regardless of the initial value. Moreover, we observe that when the measure reaches saturation, it approaches close to unity in both cases. Incase of our results, for $L = 12$, and  $q = 2$ the saturation value is $0.983$, and for $q = 3$ it is $0.998$.  While circuits composed of operators with lower entangling power require longer time scales to achieve saturation, they ultimately reach the same high value.
	\begin{table*}[t]
		\centering
		\caption{Table with all entanglement measures for the final time evolved state in  comparison to the same for the AME state.}
		\label{table_entanglement}
		\begin{tabular}{|c|c|c|c|c|c|}
			\hline
			\textbf{Measure}  & $L=8, d=2$ & $L=12, d=2$ & $L=8, d=3$ & $L=12, d=3$ & Maximum(AME) \\
			\hline
			Scott Measure& 0.933 & 0.984 & 0.987 & 0.998 & 1.0 \\
			AME-GME with GM Avg.  & 0.947 & 0.986 & 0.992 & 0.999 & 1.0 \\
			$\bar{S}_{VN}$ (Subsystem = 4) & 3.22 & - & 3.54 & - & 4 \\
			$\bar{S}_{VN}$ (Subsystem = 6) &  - & 5.27 & - & 5.54 & 6 \\
			\hline	
		\end{tabular}
	\end{table*}
	This feature of the time evolved state where the value of Scott Measure for $r = L/2$ approaches unity indicates that the state is close to an AME state. As a preliminary check we evaluate the linearised entropy $S_{L}(i)$ of the obtained state at each bond $i$, comparing it with the theoretical maximum for the resulting size of subsystems. This maximum value at bond index $i$, using the standard linearized entropy defination is given as $1 - 1/\text{min}(q^i,q^{L - i})$. The results are shown in Figure \ref{fig_entanglement_and_bonds}. A similar calculation for the AME state would lead to an overlap with the maximum bounds. While this feature of the time-evolved state is approximately similar to that of an AME state, the actual similarity between the two cannot be ascertained by this measure alone, as for the state to be near AME there must be a near maximization of multipartite entanglement. To quantify the similarity to an AME state, we consider another measure termed as the AME-GME metric \cite{ame_gme_measure}. When considering states showing many-body entanglement, we define a genuinely multipartite entangled (GME) state as one where each qubit is entangled with the remaining system. This implies that across any possible bipartition $A/\bar{A}$ the linearized entropy gives a non-zero value. The GME state is termed as an AME state, if the entanglement across all possible partitions is maximum. Taking this into account, a measure can be defined as follows \cite{ame_gme_measure}:
	\begin{equation}
		E(\psi) = \Pi_{s = 1}^{\floor{L/2}} \Pi_{k = 1}^{r_s} \frac{q^s}{q^s - 1} \Big( 1 - \tr([\rho_{k}^{s}]^2) \Big).
	\end{equation}
	Here, $r_s$ denotes the number of all possible partitions having $s$ qudits. $s$ can therefore take the values $\{1,2,3...,\floor{L/2} \}$ for a system with $L$ qudits. The density matrix $\rho_k^s$ denotes the $k$th example of a bipartition, involving $s$ parties. $r_s$ is therefore given by $^L C_s$. 
	The measure gives a non-zero value only under the condition that $\psi$ is a GME state, as the product representing $E(\psi)$ vanishes if any term of the product becomes zero corresponding to vanishing entanglement for a specific bipartition. The measure gives the maximum value of unity only in the case when $\psi$ is AME. 
	
	In its original form the measure $E(\psi)$ has the numerical value between zero and one, with a value closer to one indicating closeness to the AME state. Using the original form however, the value for our time-evolved state gives an anomalously low value around $0.1$. This can however be explained as each product term has a value $\le 1$ and $\ge 0.9$. Upon repeated multiplication this leads to the small anomalous value. To get around this, we modify the above measure by introducing a geometric mean at the end of all the products.  
	\begin{equation*}
		E_{GM}(\psi) = \Bigg( \Pi_{s = 1}^{\floor{n/2}} \Pi_{k = 1}^{r_s} \frac{q^s}{q^s - 1} \Big( 1 - \tr([\rho_{k}^{s}]^2) \Big) \Bigg)^{1/M}.
	\end{equation*}
	Here $M = \sum_{s = 1}^{\floor{L/2}} r_s$, equal to the total number of product terms. As in the other examples of multipartite measures, $E_{GM}(\psi)$ also represents an average over entanglement values upon all possible bipartitions and maybe used as a measure of multipartite entanglement. Infact, it gives a very similar variation to that of the Scott Measure for $r = \floor{L/2}$, with the added feature that its non-zero value implies the presence of genuine multipartite entanglement.
	\begin{figure*}
		\centering
		\includegraphics[scale = 0.95]{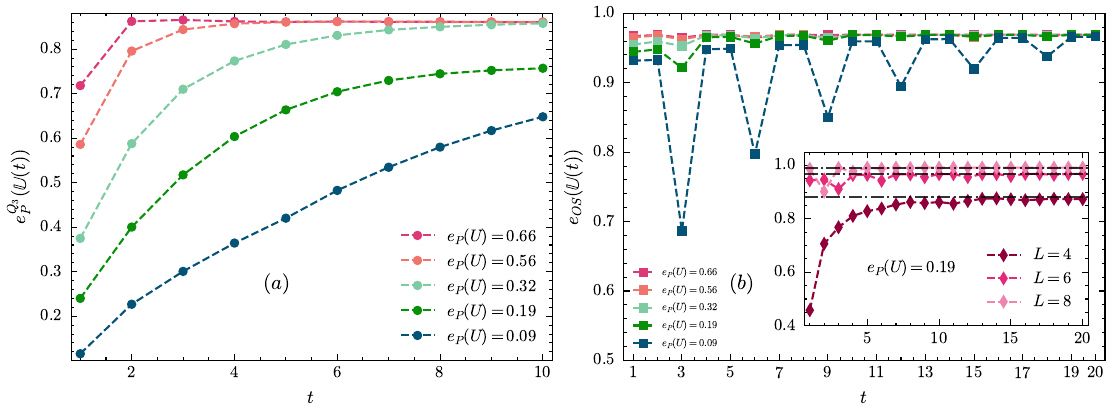}
		\caption{$(a)$ Multipartite entangling power $e_{P}^{Q_r}$, for $r = 3$, $L = 6$ and $(b)$ Operator-Space entangling power $e_{OS}$ for the entire circuit ($L = 6$) versus time steps with the single gate entangling power $e_P(\hat{U})$ as a parameter. The inset shows convergence of the metric to its maximum value for $L = 4,6 \& 8$, for a relatively small value of $e_P(\hat{U}) = 0.19$.}
		\label{fig_entangling_power_circuit}
	\end{figure*}
	
	Following the geometric-mean averaged AME-GME measure, we again find high values, being $0.986$ for $N = 12$ and $q = 2$, while reaching $0.999$ for the case when $q = 3$. Both these numbers are a multiplication result of $^{12}C_6 = 924$ different terms. In order to compare the closeness of our state with the AME State we consider yet another measure which is the average Von-Neumann entanglement entropy for all possible half system bipartitions. This measure gives the maximum value for an AME state, with the value being $L/2$, taking the base of logarithm to be the same as that of the local dimension. This measure is formally defined as:
	\begin{equation*}
		\bar{S}_{VN} = \frac{\sum_{i = 1}^{^LC_{L/2}} \Big( \sum_{j = 0}^{q^{L/2}} \lambda_j \log_q(\lambda_j) \Big)}{^LC_{L/2}} = \frac{L}{2}, \text{if}\,\, \psi = \text{AME}
	\end{equation*}
	Note that the way in which the measure is defined makes it similar to the Scott measure, based on Von-Neumann instead of linearized entropy. As shown in Table \ref{table_entanglement}, the values for all measures are close to the maximum achieved by the AME state. Here, we explain another feature (taking example of $N = 12, q = 3$), wherein the Scott($0.998/1$) and AME-GME measures ($0.999/1$) are much closer to their maximum values than the averaged Von-Neumann entropy ($5.54/6$). However, this value is an average over $q^{L/2}$ different terms, implying that each $\lambda \log_q(\lambda)$ term differs, on-average, by a small amount from the maximum. This deviation is of order $6*10^{-3}$ for $L = 12, q = 3$ and $0.049$ for $L = 8, q = 2$. The difference from maximum underlines the sensitivity of measure $\bar{S}_{VN}$ to distinguish between states having multipartite entanglement close to the maximum, and the actual AME state.  
	
	Another important feature upon observation of Table \ref{table_entanglement} is that the states furthest from AME are those corresponding to $L = 8, d = 2$ and $L = 12, d = 2$, for which an AME state does not actually exist. While it is possible to create AME states for higher $q$ using quantum circuits \cite{suhail_ame_paper_new}, our approach provides an approximate state having high multipartite entanglement which maybe useful in the absence of an actual AME state. Here it is important to stress that the high multipartite metrics arise from the convergence of the intial state to the corresponding Haar random state, following the ergodic dynamics generated by the circuit. The capability of the circuit to generate this kind of state can be quantified and is done in the following section.

	\subsection{Multipartite Entangling Power}
	\label{multipartite_entangling_powers}
	We find that the build-up of multipartite entanglement starting out for a initially separated state is dependent on the entangling powers of the constituent dual unitaries within the circuit. We now evaluate the entanglement generating capacity of the entire circuit from the multipartite perspective. For this we construct the circuits of increasing temporal depth, and check for the dependence on single gate entangling power.
	
	For a multipartite measure $E$ we can define the associated entangling power of the circuit $\mathbb{U}(t)$ as:
	\begin{equation}
		e_P^{MP}(\mathbb{U}) = \int d\mu_{n}(\psi_1,...,\psi_n) E\Big( \mathbb{U} \ket{\psi_1} \otimes...\otimes\ket{\psi_n} \Big),
	\end{equation} 
	where the measure $d\mu_n$ represents $n$ independent Haar random measures. With $E$ being the Scott measure, the multipartite entangling power $e_P^{Q_r}$ is given as \cite{Scott_Measure}:
	\begin{equation}
		e_P^{Q_r}(\mathbb{U}) = \frac{q^r}{q^r - 1}\Big( 1 - \frac{r!(r-L)!}{L!} \sum_{|S| = r} R_S(\mathbb{U})  \Big),
		\label{multi_ent_pow}
	\end{equation} 
	where $R_S(\mathbb{U})$ is given by the following expression:
	\begin{equation*}
		R_S(\mathbb{U}) = \mathcal{C}  \tr \Bigg[ \mathbb{U}^{\otimes 2} \Big( \Pi_{i = 1}^{L} P_{i,i+L} \Big) \mathbb{U}^{\dagger \otimes 2} \Big( \Pi_{i \in S} T_{i,i+L} \Big) \Bigg],
	\end{equation*}
	here $T_{ij}$ is the SWAP operator between the $i$th and $j$th factors of $(C^q)^{\otimes 2L}$, for $(0 \le i \le L)$ and $P_{ij} = (1 + T_{ij})/2$. The constant of proportionality $\mathcal{C}$ normalises the measure and is given as $\Big( 2/q(q+1) \Big)^n$.
	
	Using the working formula we calculate $e_P^{Q_r}(\mathbb{U}(t))$ over the ensemble $\{ \mathbb{U} \}_{\hat{U}'}$ for increasing value of $t$, with $e_P(\hat{U})$ appearing as a parameter (see Fig. \ref{fig_entangling_power_circuit} $(a)$). We perform calculations for the qubit case, with $L = 6$ given the computational requirements of implementing the $\mathbb{U}^{\otimes 2}$ operation. We take $r = 3$ to obtain $e_P^{Q_r}$ corresponding to the results of our previous section where $r = L/2$. We find that for all cases of $e_P(\hat{U}) > 0$ there is an increase in the multipartite entangling power $e_P^{Q_{r = 3}}(\mathbb{U}(t))$ with time $t$. With higher values of $e_P(\hat{U})$, $e_P^{Q_r}$ starts off with a high initial value at the first step, ( $0.25 \, \&  \, 0.75$ for $e_P(U) = 0.19 \, \&  \, 0.66$ respectively) and rapidly reaches maximum saturation. Even for small values of $e_P(\hat{U})$ the value of multipartite entangling power increases gradually, and it is only for SWAP that the latter is zero. The values of $e_P^{Q_{r = 3}}$ saturate at a value of around $0.9 \, \&  \, 0.7$, when $e_P(U) = 0.66 \, \&  \, 0.19$ indicating the mechanisms responsible for the delayed onset of the final state with high multipartite entanglement when $e_P(\hat{U})$ is lower, as seen in the previous section.
	
	Given the high multipartite entangling properties of the circuit $\mathbb{U}$, we consider another measure to analyse the associated dynamical nature. For this we use the recently introduced metric of operator space entangling power \cite{Zanardi_opspace_ep}. It is defined as the average operator entanglement generated by a channel given by $\mathcal{E}[\circleb] = O^{\dagger}[\circleb]O$, for an ensemble of separable haar random operators defined for a specific bipartition $(a:b)$ of the system, being acted upon by $\mathcal{E}$. It is therefore a metric defining the entanglement generating capability of $\mathbb{U}$ in the operator space, when we take $\mathcal{E}_t[\circleb] = \mathbb{U}^{\dagger}(t)[\circleb]\mathbb{U}(t)$. Following this definition the metric is given as: 
	\begin{equation}
		e_{OS}(\mathbb{U}(t)) = \mathbb{E}_{x_a,y_b} E\Big( \mathcal{E}_t(x_a \otimes y_b)\Big).
	\end{equation} 
	For our case, since the time evolution is unitary we use the notation centered on $\mathbb{U}$ instead of $\mathcal{E}$. The above form may be expressed in terms of subsystem non-commutativity measures, which results into the working expression for a half-system bipartition \cite{Zanardi_opspace_ep}:
	\begin{align*}
		e_{OS}(\mathbb{U}(t)) = 1 - \Big(1 - \frac{E(\mathbb{U})(t)}{E({\tilde{S}})}\Big)^2 - 
		\,\,\,\,\,\,\,\,\,\,\,\,\,\,\,\,\,\,\,\,\,\,\,\,\,\,\,\,\,\,\,\,\,\,\,\
		\\
		\Big(1 - \frac{E(\mathbb{U}(t)\tilde{S})}{E({\tilde{S}})}\Big)^2 - \frac{2}{q} \frac{E(\mathbb{U}(t))}{E(\tilde{S})} \frac{E(\mathbb{U}(t)\tilde{S})}{E(\tilde{S})}.
	\end{align*}
	Here $\tilde{S}$ denotes the SWAP operator exchanging the equal subsystem spaces. The value of $e_{OS}(\mathbb{U})$ for half-system bipartition extends from zero for identity operator $\mathbb{I}_{d}$ to a maximum value of $1 - 2/(d + 1)$, where $d = 2^{L}$. When $e_{OS}$ reaches its maximum value the channel $\mathcal{E}_t$ transforms every instance of the product operator $x_a \otimes y_b$ to a haar-random operator having the dimensionality of the total system. Therefore in the case of unitary evolution, $e_{OS}(\mathbb{U})$ describes the scrambling properties of the circuit operator $\mathbb{U}$.
	
	In our case, similar to $e_P^{Q_r}$, we plot $e_{OS}(\mathbb{U}(t))$ for increasing $t$ over the ensemble $\{ \mathbb{U} \}_{\hat{U}'}$ (see \ref{fig_entangling_power_circuit} $(b)$). With $L = 6$ the maximum value of $e_{OS}(\mathbb{U})$ is $0.969$, which is achieved at $t \approx 10$ even for relatively moderate values of $e_P(\hat{U})$. The metric also scales with time if $e_P(\hat{U}) > 0$. Higher values of $e_P(\hat{U})$ correspond to near maximal initial value of $e_{OS}(\mathbb{U})$. Still, $e_{OS}(\mathbb{U})$ increases with increasing time steps $t$ and actually converges to the maximum value. Time steps required for convergence sharply decreases for high $e_P(\hat{U})$ values. The convergence to maximum is more apparent for smaller $L$, which have diminished scrambling tendencies owing to the reduced number of present gates. The metric $e_{OS}(\mathbb{U})$ consequently starts out with a smaller value before gradually converging to the maximum. This convergence is expected for all values of $L$ (see \ref{fig_entangling_power_circuit}$(b)$ inset). The result further highlights that the scrambling and entanglement-generating tendencies of a circuit are related and depend on the nature of its constituent gates.
	
	\section{Relation with Spin Chains}
	\label{spin_models}
	
\begin{figure}
		\centering
		\includegraphics[scale = 0.95]{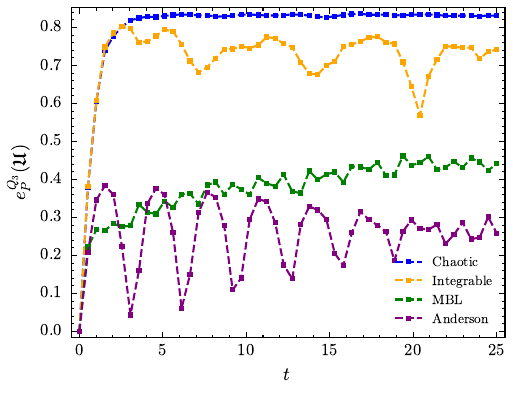}
		\caption{Variation of entangling power with time for different classes of the Transverse Field Ising Model. $L = 6$ and $r = 3$ for the measure $e_P^{Q_3}$.}
		\label{fig_entpow_ising}
	\end{figure}

	So far in this manuscript, we have considered the dynamics generated by a quantum circuit described by the evolution operator $\mathcal{U}^t$. This approach has allowed us to establish a connection between the circuit’s ergodic properties, characterized by the mixing rate, and its entanglement-generating capacity. This result remains valid for a different framework of dynamics as well, which we shall now show, taking the example of unitary dynamics generated by a physical Hamiltonian. Here in this framework we have continous time dynamics, the nature of which is determined by the interactions present in the Hamiltonian instead of the structural properties of the implemented circuit. The intrinsic nature by which the system shows ergodic properties is also very different between Hamiltonian and circuit models. For the former ergodicity is effectively classified by the scrambling properties of unitary evolution instead of decay of correlations that we have used so far for charecterising ergodic circuit dynamics. We therefore conclude our study by presenting results for different dynamical classes of the transverse field ising model with varying scrambling behaviour in Figure Fig. \ref{fig_entpow_ising}, using the multipartite entangling power metric used in the previous section. Apart from validating our results for a different dynamical setup, the obervations also indicate the usefulness of using the multipartite measure as a probe for differentiating scrambling properties within many-body systems. While it can be expected that the generated entanglement over an ensemble of initial states increases with the scrambling nature of associated time evolution, explicitly calculating the metric using equation Eq. \ref{multi_ent_pow} gives a clear separation between the different classes. 
	
	The general form of the Hamiltonian for transverse field Ising model is given as:
	\begin{equation}
		H = -\sum_{i = 1}^{L-1} \sigma^{z}_i \sigma^{z}_{i+1} - \sum_{i = 1}^{L}(h\sigma^z_{i} + g_i \sigma^x_i),
	\end{equation}
	where $h$ and $g$ represent the longitudinal and transverse Ising fields. Using these parameters we can have the Ising model corresponding to the following different dynamical classes:
	\begin{itemize}
		\item \textbf{Integrable Model}: $h = 0$ and $g_i = 1$, for all the lattice sites. This model is exactly solvable following via Jordan-Wigner mapping to a free fermionic system.
		\item \textbf{Non-Integrable (Chaotic) Model}: $h = 0.5$ and $g_i = 1.05$, $\forall i \in [1,L]$. The presence of a longitudinal field makes the hamiltonian interacting and chaotic, with the nearest neighbour spacing being described by Wigner-Dyson distribution \cite{Rohit_PRB}.
		\item \textbf{Anderson Localisation}: $h = 0$ and $g_i$ is selected randomly for each site from the uniform distribution between $[-10,10]$, this being the prototypical example for a model showing localisation. 
		\item \textbf{Many-Body Localisation}: $h = 0.5$ and $g_i$ is again randomly selected for each site using the uniform random distribution such that $g_i \in [-10,10]$. It is the interacting version of the Anderson localised model \cite{Chaos_MBL}.
	\end{itemize} 
	Having defined the models, we obtain the unitary propagator $\mathfrak{U}$ corresponding to each of these and then calculate the multipartite entangling power using equation Eq. \ref{multi_ent_pow}, for $L = 6$. We find that for the models corresponding to localisation the multipartite entangling power remains suppressed and attains a lower value compared to the integrable and chaotic examples, with the MBL phase showing still higher saturation compared to Anderson. Particularly a difference is also observed between integrable and nonintegrable (chaotic) versions not found in the operator space entangling power, owing to the requirement of local unitaries with a 2-unitary design. Such operators donot register conserved quantities in local Hilbert spaces corresponding to the integrable model. However the measure $e_P^{MP}(\mathfrak{U})$ is accurately generated by 1-unitary designs which factorise into the single particle space, differentiating the dynamics from non-integrable model \cite{Zanardi_opspace_ep}.
	
	\section{Summary and Conclusions}
	\label{summary}
	
	In this work we have considered information scrambling and entanglement generation within a finite system starting out with a generic initial state being transformed by a brickwall circuit composed of dual unitary operators. Under time evolution of this kind the space-time correlation function serves as an effective diagnostic of ergodic properties. When the correlation function is non-zero, i.e. on the lightcone connecting the space-time points, it displays constant, decaying, or oscillating behavior. The exact evaluation of the correlation function and its associated behaviour occurs in terms of a unital channel, dependent upon the individual operator $\hat{U}$ used in constructing the aforementioned circuit. This observation has prompted us to explore the role of individual operators in other dynamical aspects like scrambling and entanglement buildup.
	
	The property of individual operators most relevant to our discussion is that of entangling power. Apart from the direct relation to entanglement generation, the entangling power beyond a specific bound guarantees the circuit to be ergodically mixing. Moreover using a metric in terms of $|\lambda_1|$ to quantify this mixing we find that over several ensembles of local unitary equivalent $U$ the average mixing rate scales positively with the entangling power. In the bipartite case, we quantify entanglement buildup using the normalised entanglement measure, which we find to correlate with the mixing rate by growing monotonically with increased mixing. Taking the mixing rate allows us to consider the role of local unitaries in the dynamic properties of the circuit. In the overall picture we find that the entangling powers sets system dependent lower bounds in terms of both the ergodic nature and entanglement buildup. This relation is calculated in an exact manner for dual-unitary circuits, where the computable mixing rate provides a clear demonstration that stronger ergodicity accompanies greater entanglement generation.
	
	Moving on, we find that the entanglement structure in the time evolved state also has near maximum values for multipartite metrics of entanglement such as the Scott Measure. For circuits with $\hat{U}$ of higher entangling power the growth rate of multipartite entanglement is also higher. This implies a faster convergence to the final state with near-maximum multipartite measure. The time evolved state at saturation, with these high measures are indeed found to be close to the corresponding AME state, if it exists. In other cases we still end up with a state with high multipartite entanglement which maybe useful in the implementation of quantum computing or communication protocols.
	
	Finally we evaluate a few metrics defined for the entire circuit itself against increasing temporal depth, with the entangling power of gates being a varied parameter. The first of this is the metric of multipartite entangling power based on the Scott measure. Using this metric we observe that a circuit with greater depth has a greater capacity to generate multipartite entanglement, and that for a given depth taking individual operators of greater entangling power again enhances this capacity. To correlate this entanglement generating capability with the scrambling nature of the dynamics, we evaluate the operator space entangling power for the whole circuit, which again shows a similar qualitative behaviour. The operator space entangling power reaches its upper bounds at relatively small time steps if the entangling power is high. For both of these circuit measures there is a significant growth observed for a moderate number of time steps, except when the individual operators have vanishing entangling powers by being local unitary equivalent to the SWAP operator. 
	
	In the end, we consider different variants of the transverse field Ising model from the localising to thermal regime and based on their tendency for information scrambling, and find them to have their multipartite entangling powers in the same order as their scrambling rates. This observation serves as an example of a physical model having similar properties of our finite system under dual unitary dynamics. Here eventhough the notion of ergodicity is defined in a manner that is different from the circuit picture, a similar correlation is obtained via multipartite entanglement generation.
	
	To summarize, we conclude that the setup used in this manuscript serves as an example where the entangling nature correlates well with the ergodicity and scrambling tendencies of the system. Moreover, having examined several aspects we consider the entangling power of the individual gates forming the circuit as a strong predictor of the associated dynamical features.
	
	An interesting direction for future work is to extend our analysis of entanglement generation and scrambling beyond purely unitary circuits. This approach requires incorporating measurements within the circuit model, which is known to exhibit qualitatively richer dynamical phases \cite{claeys_measurements}, including measurement-induced transitions that sharply alter entanglement buildup and scrambling behaviour \cite{mipt_sourav}. Understanding how the entangling power of the circuits competes with or is modified by measurement processes remains an open question. A complementary avenue is to investigate ensembles of random unitaries associated with different approximate $t$-designs, which realise varying degrees of pseudorandomness and therefore distinct scrambling and entanglement-generation capabilities. Comparing the mixing rates, bipartite and multipartite entanglement growth, and operator-space diagnostics across these more general circuit families would help clarify which aspects of our present findings are universal and which rely on the special structure of dual unitaries \cite{SuzukiMitsuhashi, SuzukiEisert, RiddellKlobasBertini}. An associated line of investigation involves considering scrambling properties of heirarchical dual-unitary model \cite{hierarchical_dual}, which also lead to solvable lattice models \cite{Suhail_Membrane, solvable_lattice}. Such studies may ultimately lead to a broader classification of dynamical regimes based on entangling power and ergodic properties across both unitary and hybrid quantum circuits.
	
	\section{Acknowledgment}
	
	The authors  acknowledge  Arul Lakshminarayan for his insightful comments regarding the manuscript and  Alessandro Foligno for the calculations involving dual-unitaries in the folded tensor notation. We would like to acknowledge the support provided by the project “Study of quantum chaos and multipartite entanglement using quantum circuits” sponsored by the Anusandhan National Research Foundation (ANRF), Department of Science and Technology (DST), India, under the Core Research Grant No. CRG/2021/007095. Further, we acknowledge the computing resources of 'PARAM Shivay' at IIT(BHU) which were used in generating some results presented in this work.
	
	\section{Data Availability} 
		
		The data that support the findings of this article are openly available \cite{github_data_reference}.
	
	\bibliography{dual_dynamics}
	\appendix
	\section{Creating Examples of Dual Unitaries}
\subsection{Qubit Case}
\label{appendix_qubit_dual}
In the qubit case it is straight-forward to generate $U \in \mathcal{DU}(2)$, following the Cartan decomposition of two-qubit unitary gates \cite{Kraus_Cirac}. For a two qubit operator $A$, we have:
\begin{equation}
	A = e^{i\phi} (u_{+} \otimes u_{-}) V[J_1,J_2,J_3] (v_{+} \otimes v_{-})     
\end{equation}
where $\phi, J_1, J_2, J_3 \in \mathbb{R}$ and the local unitary operators $u_{\pm}$ and $v_{\pm}$ belong to the general SU(2) group. Here $V(J_1,J_2,J_3)$ is given as:
\begin{equation}
	V = \text{exp}[-i(J_1 \sigma_x \otimes \sigma_x + J_2 \sigma_y \otimes \sigma_y + J_3 \sigma_z \otimes \sigma_z)]
\end{equation}
Satisfying the condition for dual-unitarity leaves only a single free parameter, which then determines the value of $e_P(U)$. For $U \in \mathcal{DU}(2)$, the operator is given as:
\begin{equation}
	U[J_3] = e^{i\phi} (u_{+} \otimes u_{-}) V[\pi/4,\pi/4,J_3] (v_{+} \otimes v_{-})
\end{equation}
where we have $J_3 \in [0,\pi/4]$. This form of the two qubit gate maybe equivalently expressed as $U[J] = S_2D(J) = D(J)S_2$, where $D(J)$ is given as:
\begin{equation}
	D(J) = 
	\begin{pmatrix}
		e^{-iJ} & 0 & 0 & 0 \\
		0 & -i e^{iJ} & 0 & 0 \\
		0 & 0 & -i e^{iJ} & 0 \\
		0 & 0 & 0 & e^{-iJ} \\
	\end{pmatrix}
\end{equation}
The dual unitary operator when expressed within the Cartan form for qubits, has a very natural interpretation in terms of the Floquet transverse field ising model at the self dual point. It is given by the condition:
\begin{equation}
	|K| = |b| = \frac{\pi}{4}
\end{equation}
here, $K$ is the interaction strength, and $b$ is the periodically applied magnetic field, in a direction perpendicular to the direction of the interaction $K$.

The $J_3$ parameter appearing in the expression for the dual-unitary operator also has a direct influence in determining the entangling power of the two body operator  which can be specified as a function of $J_3$ as:
\begin{equation}
	e_{P}(U[J_3]) = \frac{2}{9}cos^2(2J_3) 
\end{equation}

\subsection{Qutrit Case}
\label{appendix_qutrit_dual}
\begin{figure}
	\centering
	\includegraphics[scale = 0.95]{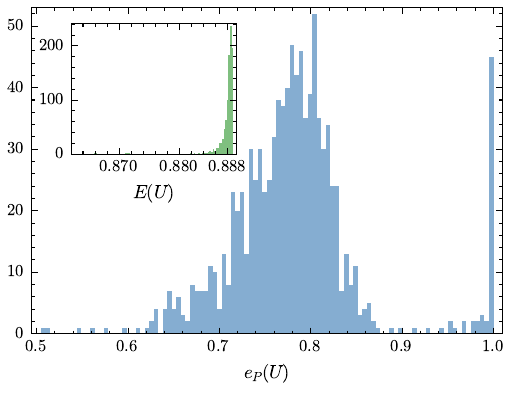}
	\caption{A distribution of entangling powers of the ensemble of operators generated by the $\mathcal{M}_R$ map, for $1000$ ensembles, each with $100$ iterations. The inset shows the distribution of operator entanglement which is saturated at the value $8/9$.}
	\label{fig_dist_ep}
\end{figure}
While the qubit case is straightforward in terms of its relation with a physical many-body hamiltonian, a similar approach is not know for the case $q = 3$. However, there exist several ways to get operators $U \in \mathcal{DU}(3)$, some of which also $\in \mathcal{U}_2(3)$. A common method relies on using permutations, which generate several dual-unitaries with different values of $e_P(U)$, including $e_P(U) = 1$ for $q = 3$. However, with this approach we do not have a continuity of successive operators as were there for the qubit case.
\begin{figure}
	\centering
	\includegraphics[scale = 0.95]{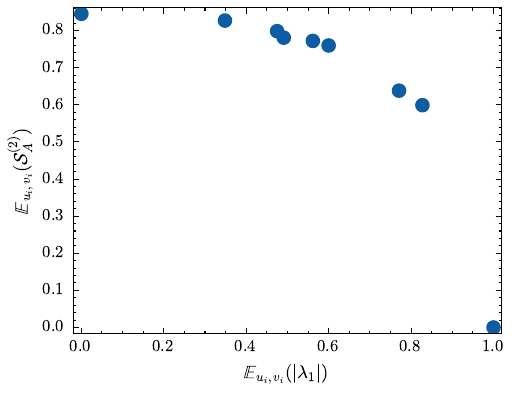}
	\caption{Average normalised entanglement $\mathcal{S}_A^{(2)}$ versus average mixing rate for an ensemble of $100$ qutrit permutation dual unitaries with increasing values of $e_P(U)$, for $t = 5$ and $\kappa_t = 2t$. The results obtained here are quite similar to those previously obtained for the class of algorithm qutrits, indicating that the source of individual operators have no effect on the variation of ergodicity and entanglement generation.}
	\label{fig_permutation_mixing}	
\end{figure}

For this reason, we employ the approach defined in Ref. \cite{Ensembles_Dual_Unitaries} where using a non-linear map applied iteratively starting out with a matrix from the Haar random distribution, generates $U \in \mathcal{DU}(q)$ with $E(U)$ arbitrarily close to maximum and a range of values for $e_P(U)$. This can even lead to $e_P(U) = 1$ for $q = 3$ and higher. The non-linear map $\mathcal{M}_R$ has the following two steps:
\begin{equation}
	\mathcal{M}_R(U):U \xrightarrow{\mathcal{R}} U^{R_2}\xrightarrow{\mathcal{PD}}V
\end{equation}
\begin{enumerate}
	\item $\mathcal{R}$ defines a linear step where $\mathcal{R}(U) = U^{R_2}$
	\item $\mathcal{PD}(U)$ is the non-linear step, where $\mathcal{PD}(U) = V$, which is the unitary closest to the operator $U^{R_2}$. The operator $V$ is found by the polar decomposition of $U^{R_2} = VP$, where $V$ is the required unitary and $P$ is a positive semi-definite Hermitian matrix. 
\end{enumerate}
In our case for generating qutrit dual unitaries we start off with an ensemble of a 1000 matrices sampled from the Haar random distribution, and iteratively apply the map $\mathcal{M}_R$ 100 times to each member of the ensemble. The resulting distribution of $E(U)$ and $e_P(U)$ is given in the figure Fig. \ref{fig_dist_ep}. Note that all the generated matrices have a value of $E(U) \approx 8/9$, while there is a range of different $e_P(U)$ values, and there are around 45 entries for $e_P(U) = 1$.

The main results of this manuscript remain unchanged by changing the source of operators $U$. To illustrate this, we shall also provide results generated for qutrit dual operators obtained using permutation matrices. Given $q = 3$ objects $\{ 1,2,3 \}$, the permutation $P$ over $q^2$ ordered pairs $\{ (i,j) \}$, for $i,j \in \{ 1,2,3 \} $ is defined as $P(i,j) = (k_{ij},l_{ij})$. Therefore taking all $q^2$ values of $i$ and $j$ the elements $k_{ij}$ and $l_{ij}$ themselves form the matrices $K$ and $L$. An example of the above maybe given as:
\begin{equation}
	\begin{pmatrix} 
		11 & 12 & 13 \\
		21 & 22 & 23 \\
		31 & 32 & 33 \\
	\end{pmatrix}
	\xrightarrow{P}
	\begin{pmatrix} 
		11 & 23 & 31 \\
		33 & 12 & 22 \\
		32 & 21 & 13 \\
	\end{pmatrix}
\end{equation}
This implies that the matrices $K$ and $L$ are:
\begin{equation}
	K = 
	\begin{pmatrix} 
		1 & 2 & 3 \\
		3 & 1 & 2 \\
		3 & 2 & 1 \\
	\end{pmatrix},
	\,\,\,
	L = 
	\begin{pmatrix} 
		1 & 3 & 1 \\
		3 & 2 & 2 \\
		2 & 1 & 3 \\
	\end{pmatrix}
\end{equation}
For a particular permutation, the qutrit operator $P$ is defined as:
\begin{equation}
	P = \sum_{i,j}^{3} \ket{k_{ij}}\ket{l_{ij}}\bra{i}\bra{j}
\end{equation}
The operator $P$ shall satisfy requirements to be dual-unitary if the matrices $K$ and $L$ have no elements repeating across the rows and columns respectively. This condition is satisfied in the example mentioned above. For the permutation to be 2-unitary, the matrices $K$ and $L$ must be Latin squares with unique entries along any row or column, and additionally must be orthogonal to generate a permutation. The resulting operator corresponds to $e_P(U) = 1$. Having a set of qutrit dual-unitaries obtained by permutations, we take the diagonalisable examples having entangling powers given by: $e_P(U) = 0.0,0.44,0.5,0.67,0.72, 0.75, 0.81, 0.89 \& 1.0$. We create an ensemble of $100$ local unitary equivalent examples for each entangling power and proceed to calculate the average entanglement velocity and mixing rate. The obtained results are shown in figure Fig. \ref{fig_permutation_mixing}.

\onecolumngrid
\section{Graphical Calculations}
\subsection{Setting}
\label{appendix_setting}
In this section we will describe the manner in which entanglement builds up for a finite system size during time evolution, following the graphical approach. To begin, let us consider a finite circuit of $N = 10$ pair products (i.e. $L = 20$) forming the initial state, which is then evolved for a single step. We then describe the density matrix $\rho_A(t = 1)$, for $A = \{ 5,6...,14\}$, which are $10$ qudits out of the total $20$. Here $i = L/4 = 5$ and $j = L - x_1 - 1 = 14.$

\begin{equation}
	\rho_A (t = 1) = \frac{1}{q^{10}} \,\,\
	\begin{tikzpicture}[baseline=(current  bounding  box.center), scale=0.25]
		
		\draw[thick] (0.5,2.5) -- (2.5,4.5);
		\draw[thick] (37.5,-0.5) -- (40.5,2.5);
		
		\draw[thick] (0.5,7.5) -- (2.5,5.5);
		\draw[thick] (37.5,10.5) -- (40.5,7.5);
		
		\draw[thick] (0.5+36,5.5) -- (4.5+36,9.5);
		\draw[thick] (0.5+36,4.5) -- (4.5+36,0.5);
		
		\draw[thick] (1.5,-0.5) -- (0.5,0.5);
		\draw[thick] (1.5,10.5) -- (0.5,9.5);
		
		\foreach \i in {0,...,8}
		{
			\draw[thick] (1.5+4*\i,-0.5) -- (6.5+4*\i,4.5);
			\draw[thick] (0.5+4*\i,5.5) -- (5.5+4*\i,10.5);
			
			\draw[thick] (0.5+4*\i,4.5) -- (5.5+4*\i,-0.5);
			\draw[thick] (1.5+4*\i,10.5) -- (6.5+4*\i,5.5);
			
		}
		\foreach \i in {0,...,9}
		{
			
			\draw[thick, fill = purple] (1.5+4*\i,-0.5) circle (0.2);
			\draw[thick, fill=purple, rounded corners=2pt] (3+4*\i, 1) rectangle (4+4*\i,2);
			\draw[thick, fill=purple, rounded corners=2pt] (1+4*\i, 3) rectangle (2+4*\i,4);
			
			\draw[thick, fill=teal, rounded corners=2pt] (1+4*\i, 6) rectangle (2+4*\i,7);
			\draw[thick, fill=teal, rounded corners=2pt] (3+4*\i, 8) rectangle (4+4*\i,9);
			\draw[thick, fill = teal] (1.5+4*\i,10.5) circle (0.2);
		}
		
		\draw[thick, dotted] (0.5,0.5) -- (4.5+36,0.5);
		\draw[thick, dotted] (0.5,2.5) -- (4.5+36,2.5);
		\draw[thick, dotted] (0.5,7.5) -- (4.5+36,7.5);
		\draw[thick, dotted] (0.5,9.5) -- (4.5+36,9.5);
		
		\foreach \i in {0,1,2}
		{
			\draw[thick] (0.5+4*\i,4.5) to[in = -110, out = 110](0.5+4*\i,5.5);
			\draw[thick] (30.5+4*\i,4.5) to[in = -70, out = 70](30.5+4*\i,5.5);
		}
		\foreach \i in {0,1}
		{
			\draw[thick] (2.5+4*\i,4.5) to[in = -70, out = 70](2.5+4*\i,5.5);
			\draw[thick] (32.5+4*\i,4.5) to[in=-110, out = 110](32.5+4*\i,5.5);
		}
		
	\end{tikzpicture}	
\end{equation}
The general factor accompanying the expression for $\rho_A(t)$ is $1/q^{N-2t}$. Using successive unitary relations, and interchanging the layers, we can simplify the figure and obtain the following expression.
\begin{equation}
	\rho_A (t = 1) = \frac{1}{q^{8}} \,\,\
	\begin{tikzpicture}[baseline=(current  bounding  box.center), scale=0.25]
		
		\draw[thick] (-0.5,0.5) to (-0.5,-0.5);
		\draw[thick] (27.5,0.5) to (27.5,-0.5);
		
		\foreach \i in {0,...,5}
		{
			\draw[thick] (-0.5+4*\i,0.5) -- (5+4*\i,6.0);
			\draw[thick] (-0.5+4*\i,-0.5) -- (5+4*\i,-6.0);
			\draw[thick] (7.5+4*\i,0.5) -- (2+4*\i,6.0);
			\draw[thick] (7.5+4*\i,-0.5) -- (2+4*\i,-6.0);
			\draw[thick, fill=purple, rounded corners=2pt] (3+4*\i, 4.0) rectangle (4+4*\i,5.0);
			\draw[thick, fill=teal, rounded corners=2pt] (3+4*\i, -4.0) rectangle (4+4*\i,-5.0);
		}
		\draw[thick] (3.5,0.5) -- (0,4.0);
		\draw[thick] (3.5,-0.5) -- (0,-4.0);
		\draw[thick] (23.5,0.5) -- (27,4.0);
		\draw[thick] (23.5,-0.5) -- (27,-4.0);
		
		\foreach \i in {0,...,6}
		{
			\draw[thick, fill=purple, rounded corners=2pt] (1+4*\i, 2.0) rectangle (2+4*\i,3.0);
			\draw[thick, fill=teal, rounded corners=2pt] (1+4*\i, -2.0) rectangle (2+4*\i,-3.0);
		}
		
		\foreach \i in {0,...,7}
		{	
			\draw[thick, fill = purple] (-0.5+4*\i,0.5) circle (0.2);
			\draw[thick, fill = teal] (-0.5+4*\i,-0.5) circle (0.2);
		}
		\draw[thick, dotted] (0,4) to[in = 120, out = -120](0,-4);
		\draw[thick, dotted] (27,4) to[in = 60, out = -60](27,-4);
		\draw[thick, dotted] (2,6) to[in = 120, out = -120](2,-6);
		\draw[thick, dotted] (25,6) to[in = 60, out = -60](25,-6);		
		
	\end{tikzpicture}	
\end{equation}
Taking powers of $\rho_A(t)$ and again applying the unitary relations gives us the aforementioned $C$ operators. For time evolution generated by $t$ time units, i.e. $2t$ layers of gates, the operator $C$ has $2t$ indices if $i$ is even and $2t+1$ indices if $i$ is odd. Note that the number of indices on one side in $C_x$ is the same as the number of involved pair products. This implies that $C_x$ maybe represented as a matrix of order $q^{x}*q^{x}$. For all remaining calculations it is the $C_x$ operator which plays the central role. For our diagram, at $t = 1$ and $i = 5$, the resulting operator is $C_3$.
\begin{equation}
	\rho_A^2(t = 1) = \frac{1}{q^{16}} \,\,\,
	\begin{tikzpicture}[baseline=(current  bounding  box.center), scale=0.25]
		
		\draw[thick] (-0.5,13.5) to (-0.5,12.5);
		\draw[thick] (-0.5,0.5) to (-0.5,-0.5);
		\draw[thick] (27.5,13.5) to (27.5,12.5);
		\draw[thick] (27.5,0.5) to (27.5,-0.5);
		
		\foreach \i in {0,...,5}
		{
			\draw[thick] (-0.5+4*\i,0.5) -- (5+4*\i,6.0);
			\draw[thick] (-0.5+4*\i,-0.5) -- (5+4*\i,-6.0);
			\draw[thick] (7.5+4*\i,0.5) -- (2+4*\i,6.0);
			\draw[thick] (7.5+4*\i,-0.5) -- (2+4*\i,-6.0);
			\draw[thick, fill=purple, rounded corners=2pt] (3+4*\i, 4.0) rectangle (4+4*\i,5.0);
			\draw[thick, fill=teal, rounded corners=2pt] (3+4*\i, -4.0) rectangle (4+4*\i,-5.0);
		}
		\draw[thick] (3.5,0.5) -- (0,4.0);
		\draw[thick] (3.5,-0.5) -- (0,-4.0);
		\draw[thick] (23.5,0.5) -- (27,4.0);
		\draw[thick] (23.5,-0.5) -- (27,-4.0);
		
		\foreach \i in {0,...,6}
		{
			\draw[thick, fill=purple, rounded corners=2pt] (1+4*\i, 2.0) rectangle (2+4*\i,3.0);
			\draw[thick, fill=teal, rounded corners=2pt] (1+4*\i, -2.0) rectangle (2+4*\i,-3.0);
		}
		
		\foreach \i in {0,...,7}
		{	
			\draw[thick, fill = purple] (-0.5+4*\i,0.5) circle (0.2);
			\draw[thick, fill = teal] (-0.5+4*\i,-0.5) circle (0.2);
		}
		\draw[thick, dotted] (0,4) to[in = 120, out = -120](0,-4);
		\draw[thick, dotted] (27,4) to[in = 60, out = -60](27,-4);
		\draw[thick, dotted] (2,6) to[in = 120, out = -120](2,-6);
		\draw[thick, dotted] (25,6) to[in = 60, out = -60](25,-6);
		
		\foreach \i in {0,...,5}
		{
			\draw[thick] (-0.5+4*\i,13.5) -- (5+4*\i,19.0);
			\draw[thick] (-0.5+4*\i,12.5) -- (5+4*\i,7.0);
			\draw[thick] (7.5+4*\i,13.5) -- (2+4*\i,19.0);
			\draw[thick] (7.5+4*\i,12.5) -- (2+4*\i,7.0);
			\draw[thick, fill=purple, rounded corners=2pt] (3+4*\i, 17.0) rectangle (4+4*\i,18.0);
			\draw[thick, fill=teal, rounded corners=2pt] (3+4*\i, 9.0) rectangle (4+4*\i,8.0);
		}
		\draw[thick] (3.5,13.5) -- (0,17.0);
		\draw[thick] (3.5,12.5) -- (0,9.0);
		\draw[thick] (23.5,13.5) -- (27,17.0);
		\draw[thick] (23.5,12.5) -- (27,9.0);
		
		\foreach \i in {0,...,6}
		{
			\draw[thick, fill=purple, rounded corners=2pt] (1+4*\i, 15.0) rectangle (2+4*\i,16.0);
			\draw[thick, fill=teal, rounded corners=2pt] (1+4*\i, 11.0) rectangle (2+4*\i,10.0);
		}
		
		\foreach \i in {0,...,7}
		{	
			\draw[thick, fill = purple] (-0.5+4*\i,13.5) circle (0.2);
			\draw[thick, fill = teal] (-0.5+4*\i,12.5) circle (0.2);
		}
		\draw[thick, dotted] (0,17) to[in = 120, out = -120](0,9);
		\draw[thick, dotted] (27,17) to[in = 60, out = -60](27,9);
		\draw[thick, dotted] (2,19) to[in = 120, out = -120](2,7);
		\draw[thick, dotted] (25, 19) to[in = 60, out = -60](25,7);	
		
		\draw (5,6) to (5,7);
		\draw (6,6) to (6,7);
		\draw (9,6) to (9,7);
		\draw (10,6) to (10,7);
		\draw (13,6) to (13,7);
		\draw (14,6) to (14,7);
		\draw (17,6) to (17,7);
		\draw (18,6) to (18,7);
		\draw (21,6) to (21,7);
		\draw (22,6) to (22,7);
		
	\end{tikzpicture}
	\,\,\, = \,\,\, \frac{1}{q^{6}} 
	\begin{tikzpicture}[baseline=(current  bounding  box.center), scale=0.25]
		
		\draw[thick] (-0.5,13.5) to (-0.5,12.5);
		\draw[thick] (-0.5,0.5) to (-0.5,-0.5);
		
		\foreach \i in {0}
		{
			\draw[thick] (-0.5+4*\i,0.5) -- (5+4*\i,6.0);
			\draw[thick] (-0.5+4*\i,-0.5) -- (5+4*\i,-6.0);
			\draw[thick] (7.5+4*\i,0.5) -- (2+4*\i,6.0);
			\draw[thick] (7.5+4*\i,-0.5) -- (2+4*\i,-6.0);
			\draw[thick, fill=purple, rounded corners=2pt] (3+4*\i, 4.0) rectangle (4+4*\i,5.0);
			\draw[thick, fill=teal, rounded corners=2pt] (3+4*\i, -4.0) rectangle (4+4*\i,-5.0);
		}
		\draw[thick] (3.5,0.5) -- (0,4.0);
		\draw[thick] (3.5,-0.5) -- (0,-4.0);
		\draw[thick] (3.5,0.5) -- (7,4.0);
		\draw[thick] (3.5,-0.5) -- (7,-4.0);
		
		\draw[thick] (7.5,0.5) -- (9,2);
		\draw[thick] (7.5,-0.5) -- (9,-2);
		\draw[thick] (7.5,13.5) -- (9,15);
		\draw[thick] (7.5,12.5) -- (9,11);
		
		\foreach \i in {0,1}
		{
			\draw[thick, fill=purple, rounded corners=2pt] (1+4*\i, 2.0) rectangle (2+4*\i,3.0);
			\draw[thick, fill=teal, rounded corners=2pt] (1+4*\i, -2.0) rectangle (2+4*\i,-3.0);
		}
		
		\foreach \i in {0,1,2}
		{	
			\draw[thick, fill = purple] (-0.5+4*\i,0.5) circle (0.2);
			\draw[thick, fill = teal] (-0.5+4*\i,-0.5) circle (0.2);
		}
		\draw[thick, dotted] (0,4) to[in = 120, out = -120](0,-4);
		\draw[thick, dotted] (2,6) to[in = 120, out = -120](2,-6);

		\foreach \i in {0}
		{
			\draw[thick] (-0.5+4*\i,13.5) -- (5+4*\i,19.0);
			\draw[thick] (-0.5+4*\i,12.5) -- (5+4*\i,7.0);
			\draw[thick] (7.5+4*\i,13.5) -- (2+4*\i,19.0);
			\draw[thick] (7.5+4*\i,12.5) -- (2+4*\i,7.0);
			\draw[thick, fill=purple, rounded corners=2pt] (3+4*\i, 17.0) rectangle (4+4*\i,18.0);
			\draw[thick, fill=teal, rounded corners=2pt] (3+4*\i, 9.0) rectangle (4+4*\i,8.0);
		}
		\draw[thick] (3.5,13.5) -- (0,17.0);
		\draw[thick] (3.5,12.5) -- (0,9.0);
		\draw[thick] (3.5,13.5) -- (7,17.0);
		\draw[thick] (3.5,12.5) -- (7,9.0);
		
		\foreach \i in {0,1}
		{
			\draw[thick, fill=purple, rounded corners=2pt] (1+4*\i, 15.0) rectangle (2+4*\i,16.0);
			\draw[thick, fill=teal, rounded corners=2pt] (1+4*\i, 11.0) rectangle (2+4*\i,10.0);
		}
		
		\foreach \i in {0,1,2}
		{	
			\draw[thick, fill = purple] (-0.5+4*\i,13.5) circle (0.2);
			\draw[thick, fill = teal] (-0.5+4*\i,12.5) circle (0.2);
		}
		\draw[thick, dotted] (0,17) to[in = 120, out = -120](0,9);
		\draw[thick, dotted] (2,19) to[in = 120, out = -120](2,7);
		
		\draw (5,6) to (5,7);
		\draw (7,9) to (7,4);
		\draw (9,11) to (9,2);
		
		\node at (9,18) {$\otimes$ 2};	
		
	\end{tikzpicture}	
\end{equation}
Since $\rho_A^{\alpha}(t)$ is described using the $C_x$ operators, the corresponding renyi entropy is given as:
\begin{equation}
	S_A^{\alpha}(t) = \frac{2}{1 - \alpha} \ln \Big( \tr[(C_{\kappa_t}C_{\kappa_t}^{\dagger})^{\alpha}] \Big)
\end{equation}
With $\kappa_t$ representing the required number of indices.
As $\tr[C_x C_x^{\dagger}] = 1$, it implies that $\sum_{i} \lambda_i = 1$, where $\lambda_i$ are the eigenvalues of $C_x C_x^{\dagger}$ matrix. Therefore the magnitude of $S_A^{\alpha}(t) \propto \sum_i \lambda_i^{\alpha}$ is also constrained between $[0,2\kappa_t\ln(q)]$. 

The lower value corresponds to $\lambda = 1$ for some $i$ whereas for the upper bound corresponds to all eigenvalues being equal. Here, $\kappa_t$ plays a significant role in determining the limits of $S_A^{\alpha}(t)$ as the dimensions of $C_{\kappa_t}$ operator are given by $q^{\kappa_t}$, which also specifies the number of eigenvalues. Thus, by properties of the $C$ operator, $S_A^{\alpha}(t) \le 2\kappa_t\ln(q)$. The maximal value is obtained for the matrix $m$ being unitary, and this being the maximal velocity is the upper bound for all orders of Renyi entropy, which converge for the maximal case \cite{Sarang_ent_val, Dual_Dynamics_OTOC}. The entanglement velocity $v_E^{\alpha}(t)$ at time $t$ can therefore be given as the ratio of $S_A^{(\alpha)}(t)/2\kappa_t\ln(q)$.

For $\alpha = 2$, we have $S_A^{2}(t) = -2\ln\Big( \tr[(C_{\kappa_t}C_{\kappa_t}^{\dagger})^2] \Big)$. For creating the ensembles of different $U'$ we select the local unitaries $\{ u_1,u_2,v_1,v_2 \}$ from the Haar random distribution. Once the ensemble is created, we construct the $C'$ operator for each $U'$ within the ensemble to evaluate the average for the Renyi entropy. 

\subsection{Lower Bounds for Entanglement Growth}
\label{appendix_entanglement_velocity}

Calculating the ensemble average of Renyi entropy implies averaging over the scalar quantity having the form $-2\ln(\tr[({C'}_{x}{C'}_{x}^{\dagger})^2] )$ for all members of the ensemble. A particular example of the ${C'}_x$ operator for $x = 4$, involving a specific $U'$ is given as follows. Note that for each $C'$, the initial state is kept fixed.
\begin{equation}
	{C'}_{x = 4} = \frac{1}{q^2} \,\,\,
	\begin{tikzpicture}[baseline=(current  bounding  box.center), scale=0.35]
		
		\foreach \i in {0,1,2,3}
		{
			
			\draw[thick] (-0.5+4*\i,0.5) -- (7+2*\i,8-2*\i);
			\draw[thick] (11.5-4*\i,0.5) -- (4-2*\i,8-2*\i);
		}
		\foreach \i in {0,1,2}
		{
			\draw[thick, fill=gray, rounded corners=2pt] (1+4*\i, 2.0) rectangle (2+4*\i,3.0);
		} 
		
		\foreach \i in {0,1}
		{
			\draw[thick, fill=gray, rounded corners=2pt] (3+4*\i, 4.0) rectangle (4+4*\i,5.0);
		} 
		\foreach \i in {0}
		{
			\draw[thick, fill=gray, rounded corners=2pt] (5+4*\i, 6.0) rectangle (6+4*\i,7.0);
		}
		\node at (14,4) {;};
		\foreach \i in {0,1,2,3}
		{
			\draw[thick, fill = gray] (-0.5+4*\i,0.5) circle (0.2);
		}

	\end{tikzpicture}
	\,\,\,
	\begin{tikzpicture}[baseline=(current  bounding  box.center), scale=0.5]
		\draw[thick] (0.5,0.5) -- (2.5,2.5);
		\draw[thick] (2.5,0.5) -- (0.5,2.5);
		\draw[thick, fill=gray, rounded corners=2pt] (1,1) rectangle (2,2);
		\node at (3.5,1.5) {$=$};
		\draw[thick] (4.5,0.5) -- (6.5,2.5);
		\draw[thick] (6.5,0.5) -- (4.5,2.5);
		\draw[thick, fill=purple, rounded corners=2pt] (5,1) rectangle (6,2);
		\draw[thick, fill = green] (4.5,0.5) circle (0.3);
		\draw[thick, fill = yellow] (6.5,2.5) circle (0.3);
		\draw[thick, fill = blue] (6.5,0.5) circle (0.3);
		\draw[thick, fill = black] (4.5,2.5) circle (0.3);
		\node at (5.5,0.5) {$u_1$};
		\node at (7.5,2.5) {$v_2$};
		\node at (7.5,0.5) {$u_2$};
		\node at (5.5,2.5) {$v_1$};
		
	\end{tikzpicture}
\end{equation}
Therefore in this diagram at each space-time point there is a local unitary operator. As a feature of the arrangement that we have used, the random operators at any two sites at a distance of two lattice constants are exactly the same, provided that they are not at the outermost edges of the diagram. This occurs as the local unitaries $\{ u_1,u_2,v_1,v_2 \}$ after being drawn independently, are repeated throughout the circuit, being  mounted the operator $U \in \mathcal{DU}(q)$. This arrangement is therefore a special case of the setup used in \cite{Dual_Dynamics_Foligno}, and hence we can conclude that for the infinite case, using operators $U'$ shall also give the same results. We use the similar approach in terms of simplifying the operator $\mathbb{E}_{u,v}\big[\tr(({C'}_{x}{C'}_{x}^{\dagger})^2)\big]$ using the folded tensor notation.

We define the term $\mathcal{K}_x$ as $\tr([{C'}_{x}{C'}_{x}^{\dagger}]^2)$. 
Following ensemble averaging, the average entanglement entropy is given by $\bar{S}_A^2(t_x) = \mathbb{E} (-2\ln(\mathcal{K}_x))$. The time $t_x$ corresponds to the number of indices being $x$, i.e. $\kappa_{t_x} = x$. With $f(x) = -2\ln(x)$ being a convex function, using Jensen's inequality $\mathbb{E}(f(x)) \le f(\mathbb{E}(x))$ we have $\bar{S}_A^2(t_x) = \mathbb{E}(-2\ln(\mathcal{K}_x)) \ge -2\ln(\mathbb{E}(\mathcal{K}_x)) $. Thus evaluating $\mathbb{E}(\mathcal{K}_x)$ and taking the logarithm and adding the multiplicative factor $-2$ gives us a lower bound of the average entanglement entropy $\bar{S}_A^2(t_x)$.

This quantity $\mathbb{E}(\mathcal{K}_x)$ is denoted in the main text as $\mathcal{P}_x$. To evaluate the expression for $\mathcal{P}_x$  we use the folded tensor notation, similar to \cite{Dual_Dynamics_Foligno}:

\begin{equation}
	\mathcal{P}_{x = 4} = \frac{1}{q^2} \,\,\,
	\begin{tikzpicture}[baseline=(current  bounding  box.center), scale=0.35]
		
		\foreach \i in {0,1,2,3}
		{
			
			\draw[thick] (-0.5+4*\i,0.5) -- (7+2*\i,8-2*\i);
			\draw[thick] (11.5-4*\i,0.5) -- (4-2*\i,8-2*\i);
			\draw[thick, fill = green] (-0.5+4*\i,0.5) circle (0.2);
		}
		
		\foreach \i in {0,1,2}
		{
			\draw[thick, fill=green, rounded corners=2pt] (1+4*\i, 2.0) rectangle (2+4*\i,3.0);
		} 
		
		\foreach \i in {0,1}
		{
			\draw[thick, fill=green, rounded corners=2pt] (3+4*\i, 4.0) rectangle (4+4*\i,5.0);
		} 
		\foreach \i in {0}
		{
			\draw[thick, fill=green, rounded corners=2pt] (5+4*\i, 6.0) rectangle (6+4*\i,7.0);
		}
		\foreach \i in {0,1,2,3}
		{	 
			\draw[thick, fill = white] (4-2*\i,8-2*\i) circle (0.3); 
		}
		\foreach \i in {0,1,2,3}
		{	 
			\draw[thick, fill = white] (7+2*\i,8-2*\i) rectangle (7.5+2*\i,8.5-2*\i); 
		}	
		\foreach \i in {0,1,2,3}
		{
			\draw[thick, fill = green] (-0.5+4*\i,0.5) circle (0.2);
		}
		
		\node at (14,4) {;};
		
	\end{tikzpicture}
	\,\,\,
	\begin{tikzpicture}[baseline=(current  bounding  box.center), scale=0.5]
		\draw[thick] (0.5,0.5) -- (2.5,2.5);
		\draw[thick] (2.5,0.5) -- (0.5,2.5);
		\draw[thick, fill=green, rounded corners=2pt] (1,1) rectangle (2,2);
		\node at (5.0,2.5) {$= U' \otimes_r U'$};
		\node at (7.5,0.5) {$= (P \otimes_r P) (U \otimes_r U) (P \otimes_r P)$};

	\end{tikzpicture}
	\label{haar_projector}
\end{equation}
with $\ket{\circlew} = \frac{1}{d}\sum \delta_{ij} \delta_{kl} \ket{ijkl}$ and $\ket{\squarew} = \frac{1}{d} \sum \delta_{il} \delta_{jk} \ket{ijkl}$. 

Upon folding up the tensor diagram, the local unitaries and their conjugates end up twice stacked on top of each other, and following ensemble averaging lead to the projectors acting on the space spanned by $\ket{\circlew}$ and $\ket{\squarew}$. The exact form of the ensemble average leading to the projectors can be obtained for several cases \cite{Zanardi_Entangling_Power} using the group theoretic approach involving the Schur's lemma, or equivalently by a linear algebra approach \cite{Tutorial_Haar}. Note that the above equation Eq. \ref{haar_projector} is exactly true only when there are independently sampled Haar random unitaries at each space-time point of the $1+1D$ lattice, also known as the inhomogeneous circuit model \cite{Aravinda_Sir_PRR}. In our case, $\mathbb{U}(t) =  \mathcal{U}^t$ which is essentially a Floquet type setup, the equation is only approximately true. However, for the finite system size we find that both inhomogeneous and Floquet type circuits generate exactly same averages, implying that the approximation is indeed justified as further evidenced in references \cite{Bhargavi_Jonnadula1, Dual_Dynamics_Foligno}.

A gist of the approach following the Schur's lemma is as follows: The lemma states that if $M(g),\ \forall g \in G$ is an irreducible representation for the group $G$ on the vector space $V$, and there $\exists A$, such that $AM(g) = M(g)A$, then $A \propto \mathbb{I}$ for the vector space $V$. Here $G$ is $\mathcal{U}(q)$ and $M(g) \in U^{\otimes 4}(q)$ with $M(g) = \{ U \otimes U^{*} \otimes U \otimes U^{*}, \forall U \in \mathcal{U}(q) \}$. The matrix $A$ corresponds to $\mathbb{E}(w\otimes w^{*}\otimes w \otimes w^{*})$, where $w \in$ haar random distribution over $\mathcal{U}(q)$. It is local unitary invariant $\forall U \in \mathcal{U}(q)$ and thus commutes with $M(g), \,\, \forall g \in G$. This implies that $A \propto \mathbb{I}$ acts on the vector space $V$ and is given by the projector from the usual computational basis to $V$, depending on symmetry considerations.

For example, in our case for $V$ upon which $M(g)$ is described is invariant under $S_{12|34}$ and $S_{13|24}$ where $S_{ab|cd}$ is the SWAP operator exchanging $ab$ with $cd$. Thus $V$ is spanned by $\ket{\circlew}$ and $\ket{\squarew}$, which satisfy the same symmetries. The states $\ket{\circlew}$ and $\ket{\squarew}$ are not orthogonal as $\bra{\squarew}\circlew\rangle = 1/d$. Starting from either and applying the Gram-Schmidt orthogonalization process, we get a linearly independent basis. This is represented as $\{ \ket{\circlew},\ket{\circleb} \}$ or $\{ \ket{\squarew},\ket{\squareb} \}$ where:
\begin{align}
	\ket{\circleb} = \frac{q\ket{\squarew} - \ket{\circlew}}{\sqrt{q^2 - 1}} \\
	\ket{\squareb} = \frac{q\ket{\circlew} - \ket{\squareb}}{\sqrt{q^2 - 1}}
	\label{S1}
\end{align}
The projector $P$ can then be given as:
\begin{equation}
	P = \ket{\circlew}\bra{\circlew} + \ket{\circleb}\bra{\circleb} = \ket{\squarew}\bra{\squarew} + \ket{\squareb}\bra{\squareb} 
\end{equation}
Using this expression simplifies $\mathcal{P}_x$, as the projector $P$ acting on the states  $\ket{\circlew}$ and $\ket{\squarew}$ is straightforward. Further, the property of dual-unitarity comes in handy due to which we have:

\begin{equation}
	\begin{tikzpicture}[baseline=(current  bounding  box.center), scale=0.5]
		\draw[thick] (1,1) -- (3,3);
		\draw[thick] (2,1) -- (0,3);
		\draw[thick, fill = green, rounded corners=2pt] (1,1) rectangle (2,2);
		\draw[thick, fill = white] (3,3) circle (0.3);
		\draw[thick, fill = white] (0,3) circle (0.3);
		\node at (4.0,1.5) {$=$};
		\draw[thick] (6,0) -- (6,3);
		\draw[thick] (7,0) -- (7,3);
		\draw[thick, fill = white] (6,3) circle (0.3);
		\draw[thick, fill = white] (7,3) circle (0.3);
		\node at (8.0,1.5) {;};
		\draw[thick] (9,1) -- (11,3);
		\draw[thick] (9,2) -- (11,0);
		\draw[thick, fill = green, rounded corners=2pt] (9,1) rectangle (10,2);
		\draw[thick, fill = white] (11,3) circle (0.3);
		\draw[thick, fill = white] (11,0) circle (0.3);
		\node at (13.0,1.5) {$=$};
		\draw[thick] (14,2) -- (17,2);
		\draw[thick] (14,1) -- (17,1);
		\draw[thick, fill = white] (17,2) circle (0.3);
		\draw[thick, fill = white] (17,1) circle (0.3);
	\end{tikzpicture}	
	\label{S2}
\end{equation}

\begin{equation}
	\begin{tikzpicture}[baseline=(current  bounding  box.center), scale=0.5]
		\draw[thick] (1,1) -- (3,3);
		\draw[thick] (2,1) -- (0,3);
		\draw[thick, fill = green, rounded corners=2pt] (1,1) rectangle (2,2);
		\draw[thick, fill = white] (2.75,2.75) rectangle (3.25,3.25);
		\draw[thick, fill = white] (-0.25,2.75) rectangle (0.25,3.25);
		\node at (4.0,1.5) {$=$};
		\draw[thick] (6,0) -- (6,3);
		\draw[thick] (7,0) -- (7,3);	
		\draw[thick, fill = white] (5.75,2.75) rectangle (6.25,3.25);
		\draw[thick, fill = white] (6.75,2.75) rectangle (7.25,3.25);
		\node at (8.0,1.5) {;};
		\draw[thick] (9,1) -- (11,3);
		\draw[thick] (9,2) -- (11,0);
		\draw[thick, fill = green, rounded corners=2pt] (9,1) rectangle (10,2);
		\draw[thick, fill = white] (10.75,2.75) rectangle (11.25,3.25);
		\draw[thick, fill = white] (10.75,-0.25) rectangle (11.25,0.25);
		\node at (13.0,1.5) {$=$};
		\draw[thick] (14,2.5) -- (17,2.5);
		\draw[thick] (14,0.5) -- (17,0.5);
		\draw[thick, fill = white] (16.75,2.25) rectangle (17.25,2.75);
		\draw[thick, fill = white] (16.75,0.25) rectangle (17.25,0.75);
	\end{tikzpicture}
	\label{S3}	
\end{equation}
Further we also have:
\begin{equation}
	\bra{\squarew}(m\otimes_r m^{*})^{\otimes_r 2}\ket{\circlew} = \,\,\,
	\begin{tikzpicture}[baseline=(current  bounding  box.center), scale=0.5]
		\draw[thick] (0,0) -- (1,1);
		\draw[thick] (0,0) -- (-1,1);
		\draw[thick, fill = white] (1,1) circle (0.2);
		\draw[thick, fill = white] (-1.25,0.75) rectangle (-0.75,1.25);
		\draw[thick, fill = green] (0,0) circle (0.3); 				
	\end{tikzpicture}
	\,\,\,
	= \frac{\tr((mm^{\dagger})^2)}{q^2} = \frac{c}{q}
	\label{cd_equation}
\end{equation}
Here $c = \frac{tr((mm^{\dagger}))^2}{q}$ appears as a parameter which characterises the random initial state. Now in the right-most and left-most corner of the tensor $\mathcal{P}_x$, owing to the presence of $P$ we have:

\begin{equation}
	\bra{P}m\ket{\circlew} = \ket{\circlew}\bra{\circlew}m\ket{\circlew} + \ket{\circleb}\bra{\circleb}m\ket{\circlew} = \ket{\circlew} + \ket{\circleb}\Big( \frac{\bra{\squarew}d - \bra{\circlew}}{\sqrt{q^2 - 1}} m \ket{\circlew} \Big) 
\end{equation}

\begin{equation}
	\implies \bra{P}m\ket{\circlew} = \ket{\circlew} + \frac{c - 1}{\sqrt{q^2 - 1}}\ket{\circleb} = \,\,\,
	\begin{tikzpicture}[baseline=(current  bounding  box.center), scale=0.5]
		\draw[thick] (0,0) -- (1,1);
		\draw[thick] (0,0) -- (-1,1);
		\draw[thick, fill = white] (1,1) circle (0.2);
		\node at (2,1) {$+$};
	\end{tikzpicture}
	\frac{c - 1}{\sqrt{q^2 - 1}} \,\,\,
	\begin{tikzpicture}[baseline=(current  bounding  box.center), scale=0.5]
		\draw[thick] (0,0) -- (1,1);
		\draw[thick] (0,0) -- (-1,1);
		\draw[thick, fill = black] (1,1) circle (0.2);
	\end{tikzpicture}
	\label{S4}
\end{equation}	

\begin{equation}
	\bra{P}m\ket{\squarew} = \ket{\squarew}\bra{\squarew}m\ket{\squarew} + \ket{\squareb}\bra{\squareb}m\ket{\squareb} = \ket{\squarew} + \ket{\squareb}\Big( \frac{\bra{\circlew}d - \bra{\squarew}}{\sqrt{q^2 - 1}} m \ket{\squarew} \Big) 
\end{equation}

\begin{equation}
	\implies \bra{P}m\ket{\squarew} = \ket{\squarew} + \frac{c - 1}{\sqrt{q^2 - 1}}\ket{\squareb} = \,\,\,
	\begin{tikzpicture}[baseline=(current  bounding  box.center), scale=0.5]
		\draw[thick] (0,0) -- (1,1);
		\draw[thick] (0,0) -- (-1,1);
		\draw[thick, fill = white] (0.75,0.75) rectangle (1.25,1.25);
		\node at (2,1) {$+$};
	\end{tikzpicture}
	\frac{c - 1}{\sqrt{q^2 - 1}} \,\,\,
	\begin{tikzpicture}[baseline=(current  bounding  box.center), scale=0.5]
		\draw[thick] (0,0) -- (1,1);
		\draw[thick] (0,0) -- (-1,1);
		\draw[thick, fill = black] (0.75,0.75) rectangle (1.25,1.25);
	\end{tikzpicture}
	\label{S5}
\end{equation}	
We denote the term $\frac{c-1}{\sqrt{q^2 - 1}}$ as $\chi$. As $c \in [1,q]$ it implies that the value of $\chi$ is always less than unity as $\chi \in [0,\sqrt{\frac{q-1}{q+1}}]$. Our approach is to observe the nature of $\tr(\mathcal{P}_x)$ for increasing values of $x$, and deduce the nature of the entanglement buildup and the effect of parameters such as $c$ and $e_P(U)$. In doing so, the successive calculations make use of the previous results, and we get a nearly explicit form for $\tr(\mathcal{P}_x)$. The smallest term corresponding to $\mathcal{P}_x$ is trivial expression for $\mathcal{P}_1 = \bra{\circlew}m\ket{\squareb} = c/d$. The next term is $\mathcal{P}_2$ which is given by the following expression, making use of Eq. \ref{S4} and Eq. \ref{S5}:
\begin{equation}
	\begin{tikzpicture}[baseline=(current  bounding  box.center), scale=0.5]
		\draw[thick] (0,0) -- (3,3);
		\draw[thick] (3,0) -- (0,3);
		\draw[thick] (0,0) -- (-1,1);
		\draw[thick] (3,0) -- (4,1);
		\draw[thick, fill = green] (0,0) circle (0.3);
		\draw[thick, fill = green] (3,0) circle (0.3);
		\draw[thick, fill = green, rounded corners=2pt] (1,1) rectangle (2,2);
		\draw[thick, fill = white] (4,1) circle (0.3);
		\draw[thick, fill = white] (3,3) circle (0.3);
		\draw[thick, fill = white] (-0.25,2.75) rectangle (0.25,3.25);
		\draw[thick, fill = white] (-1.25,0.75) rectangle (-0.75,1.25);
	\end{tikzpicture}
	\,\,\, = \,\,\,
	\begin{tikzpicture}[baseline=(current  bounding  box.center), scale=0.5]
		\draw[thick] (0,0) -- (3,3);
		\draw[thick] (3,0) -- (0,3);
		\draw[thick, fill = white] (3,0) circle (0.3);
		\draw[thick, fill = green, rounded corners=2pt] (1,1) rectangle (2,2);
		\draw[thick, fill = white] (3,3) circle (0.3);
		\draw[thick, fill = white] (-0.25,2.75) rectangle (0.25,3.25);
		\draw[thick, fill = white] (-0.25,-0.25) rectangle (0.25,0.25);	
	\end{tikzpicture}
	\,\,\, + \,\,\, \chi
	\begin{tikzpicture}[baseline=(current  bounding  box.center), scale=0.5]
		\draw[thick] (0,0) -- (3,3);
		\draw[thick] (3,0) -- (0,3);
		\draw[thick, fill = black] (3,0) circle (0.3);
		\draw[thick, fill = green, rounded corners=2pt] (1,1) rectangle (2,2);
		\draw[thick, fill = white] (3,3) circle (0.3);
		\draw[thick, fill = white] (-0.25,2.75) rectangle (0.25,3.25);
		\draw[thick, fill = white] (-0.25,-0.25) rectangle (0.25,0.25);	
	\end{tikzpicture}
	\,\,\, + \,\,\, \chi
	\begin{tikzpicture}[baseline=(current  bounding  box.center), scale=0.5]
		\draw[thick] (0,0) -- (3,3);
		\draw[thick] (3,0) -- (0,3);
		\draw[thick, fill = white] (3,0) circle (0.3);
		\draw[thick, fill = green, rounded corners=2pt] (1,1) rectangle (2,2);
		\draw[thick, fill = white] (3,3) circle (0.3);
		\draw[thick, fill = white] (-0.25,2.75) rectangle (0.25,3.25);
		\draw[thick, fill = black] (-0.25,-0.25) rectangle (0.25,0.25);	
	\end{tikzpicture}
	\,\,\, + \,\,\, \chi^2
	\begin{tikzpicture}[baseline=(current  bounding  box.center), scale=0.5]
		\draw[thick] (0,0) -- (3,3);
		\draw[thick] (3,0) -- (0,3);
		\draw[thick, fill = black] (3,0) circle (0.3);
		\draw[thick, fill = green, rounded corners=2pt] (1,1) rectangle (2,2);
		\draw[thick, fill = white] (3,3) circle (0.3);
		\draw[thick, fill = white] (-0.25,2.75) rectangle (0.25,3.25);
		\draw[thick, fill = black] (-0.25,-0.25) rectangle (0.25,0.25);	
	\end{tikzpicture}
	\label{P2_equation}
\end{equation}
using relations Eq. \ref{S2} and Eq. \ref{S3}, we have: 
\begin{equation}
	\implies \mathcal{P}_2 = \,\,\, \Big(
	\begin{tikzpicture}[baseline=(current  bounding  box.center), scale=0.5]
		\draw[thick] (0,0) -- (1,0);
		\draw[thick, fill = white] (0,0) circle (0.2);
		\draw[thick, fill = white] (0.75,-0.25) rectangle (1.25,0.25);
	\end{tikzpicture}
	\Big)^2 + \,\,\, \chi \Bigg(
	\begin{tikzpicture}[baseline=(current  bounding  box.center), scale=0.5]
		\draw[thick] (0,0.5) -- (1,0.5);
		\draw[thick, fill = white] (0,0.5) circle (0.2);
		\draw[thick, fill = white] (0.75,0.25) rectangle (1.25,0.75);
		
		\draw[thick] (0,-0.5) -- (1,-0.5);
		\draw[thick, fill = black] (0,-0.5) circle (0.2);
		\draw[thick, fill = white] (0.75,-0.25) rectangle (1.25,-0.75);
	\end{tikzpicture}
	\Bigg) + \,\,\, \chi \Bigg(
	\begin{tikzpicture}[baseline=(current  bounding  box.center), scale=0.5]
		\draw[thick] (0,0.5) -- (1,0.5);
		\draw[thick, fill = white] (0,0.5) circle (0.2);
		\draw[thick, fill = white] (0.75,0.25) rectangle (1.25,0.75);
		
		\draw[thick] (0,-0.5) -- (1,-0.5);
		\draw[thick, fill = white] (0,-0.5) circle (0.2);
		\draw[thick, fill = black] (0.75,-0.25) rectangle (1.25,-0.75);
	\end{tikzpicture}
	\Bigg) \,\,\, + \,\,\, \chi^2
	\begin{tikzpicture}[baseline=(current  bounding  box.center), scale=0.5]
		\draw[thick] (0,0) -- (3,3);
		\draw[thick] (3,0) -- (0,3);
		\draw[thick, fill = black] (3,0) circle (0.3);
		\draw[thick, fill = green, rounded corners=2pt] (1,1) rectangle (2,2);
		\draw[thick, fill = white] (3,3) circle (0.3);
		\draw[thick, fill = white] (-0.25,2.75) rectangle (0.25,3.25);
		\draw[thick, fill = black] (-0.25,-0.25) rectangle (0.25,0.25);	
	\end{tikzpicture}
	\label{P2_equation_simplified1}
\end{equation}
now using relations Eq. \ref{S1}, we can simplify the final term:
\begin{equation}
	\begin{tikzpicture}[baseline=(current  bounding  box.center), scale=0.5]
		\draw[thick] (0,0) -- (3,3);
		\draw[thick] (3,0) -- (0,3);
		\draw[thick, fill = black] (3,0) circle (0.3);
		\draw[thick, fill = green, rounded corners=2pt] (1,1) rectangle (2,2);
		\draw[thick, fill = white] (3,3) circle (0.3);
		\draw[thick, fill = white] (-0.25,2.75) rectangle (0.25,3.25);
		\draw[thick, fill = black] (-0.25,-0.25) rectangle (0.25,0.25);	
	\end{tikzpicture}
	\,\,\, = \,\,\, \frac{1}{q^2 - 1} \Biggr[ \,\,\, q^2\,\,\,
	\begin{tikzpicture}[baseline=(current  bounding  box.center), scale=0.5]
		\draw[thick] (0,0) -- (3,3);
		\draw[thick] (3,0) -- (0,3);
		\draw[thick, fill = white] (0,0) circle (0.3);
		\draw[thick, fill = green, rounded corners=2pt] (1,1) rectangle (2,2);
		\draw[thick, fill = white] (3,3) circle (0.3);
		\draw[thick, fill = white] (-0.25,2.75) rectangle (0.25,3.25);
		\draw[thick, fill = white] (2.75,-0.25) rectangle (3.25,0.25);	
	\end{tikzpicture}
	\,\,\, -q
	\begin{tikzpicture}[baseline=(current  bounding  box.center), scale=0.5]
		\draw[thick] (0,0) -- (3,3);
		\draw[thick] (3,0) -- (0,3);
		\draw[thick, fill = white] (3,0) circle (0.3);
		\draw[thick, fill = green, rounded corners=2pt] (1,1) rectangle (2,2);
		\draw[thick, fill = white] (3,3) circle (0.3);
		\draw[thick, fill = white] (-0.25,2.75) rectangle (0.25,3.25);
		\draw[thick, fill = white] (0,0) circle (0.3);	
	\end{tikzpicture}
	\,\,\, -q
	\begin{tikzpicture}[baseline=(current  bounding  box.center), scale=0.5]
		\draw[thick] (0,0) -- (3,3);
		\draw[thick] (3,0) -- (0,3);
		\draw[thick, fill = green, rounded corners=2pt] (1,1) rectangle (2,2);
		\draw[thick, fill = white] (3,3) circle (0.3);
		\draw[thick, fill = white] (-0.25,2.75) rectangle (0.25,3.25);
		\draw[thick, fill = white] (2.75,-0.25) rectangle (3.25,0.25);
		\draw[thick, fill = white] (-0.25,-0.25) rectangle (0.25,0.25);	
	\end{tikzpicture}
	\,\,\, + \,\,\,
	\begin{tikzpicture}[baseline=(current  bounding  box.center), scale=0.5]
		\draw[thick] (0,0) -- (3,3);
		\draw[thick] (3,0) -- (0,3);
		\draw[thick, fill = white] (3,0) circle (0.3);
		\draw[thick, fill = green, rounded corners=2pt] (1,1) rectangle (2,2);
		\draw[thick, fill = white] (3,3) circle (0.3);
		\draw[thick, fill = white] (-0.25,2.75) rectangle (0.25,3.25);
		\draw[thick, fill = white] (-0.25,-0.25) rectangle (0.25,0.25);	
	\end{tikzpicture}
	\,\,\,\Biggr]
	\label{P2_intermediate}
\end{equation}
In this expression all terms except the first can be reduced by the repeated application of the relations given in Eq. \ref{S2} and \ref{S3}. The first term is obtained using the definition of entangling power using the $T_2$ transformation. 
\begin{equation}
	e_P(U) = \frac{q^4 - \tr[(\tilde{U}^{T_2}(\tilde{U}^{T_2})^{\dagger})^2]}{q^4 - q^2}
\end{equation}
\begin{equation*}
	\implies \frac{\tr[(\tilde{U}^{T_2}(\tilde{U}^{T_2})^{\dagger})^2]}{q^4} = 
	\begin{tikzpicture}[baseline=(current  bounding  box.center), scale=0.5]
		\draw[thick] (0,0) -- (3,3);
		\draw[thick] (3,0) -- (0,3);
		\draw[thick, fill = white] (0,0) circle (0.3);
		\draw[thick, fill = green, rounded corners=2pt] (1,1) rectangle (2,2);
		\draw[thick, fill = white] (3,3) circle (0.3);
		\draw[thick, fill = white] (-0.25,2.75) rectangle (0.25,3.25);
		\draw[thick, fill = white] (2.75,-0.25) rectangle (3.25,0.25);	
	\end{tikzpicture}
	= 1 -e_p + \frac{e_p}{q^2}
\end{equation*}
Hence, we may now simplify the relation given by Eq. \ref{P2_intermediate}.
\begin{equation}
	\begin{tikzpicture}[baseline=(current  bounding  box.center), scale=0.5]
		\draw[thick] (0,0) -- (3,3);
		\draw[thick] (3,0) -- (0,3);
		\draw[thick, fill = black] (3,0) circle (0.3);
		\draw[thick, fill = green, rounded corners=2pt] (1,1) rectangle (2,2);
		\draw[thick, fill = white] (3,3) circle (0.3);
		\draw[thick, fill = white] (-0.25,2.75) rectangle (0.25,3.25);
		\draw[thick, fill = black] (-0.25,-0.25) rectangle (0.25,0.25);	
	\end{tikzpicture}
	\,\,\, = 
	\frac{1}{q^2-1} \Bigg[ q^2\Big(1 - e_P + \frac{e_P}{q^2}\Big) -1 -1 -\frac{1}{q^2} \Bigg] = 
	1 - e_P - \frac{1}{q^2}	
\end{equation}
Collecting all the terms in Eq. \ref{P2_equation}, we get:
\begin{equation}
	P_2 = \frac{1}{q^2} + 2\chi \Big( \frac{1}{q^2} \sqrt{q^2 - 1} \Big) + \chi^2 \Big( 1 - e_P -\frac{1}{q^2} \Big)
\end{equation}

The next term for $\mathcal{P}_3$ represents the first non-trivial term of the series as it cannot be exactly evaluated and requires an inequality to constraint its values. The tensor corresponding to $P_3$ is written down, and simplified along the lines of Eq. \ref{P2_equation}, replacing the initial state tensor in the right-most and left-most corner following equations Eq. \ref{S4} and Eq. \ref{S5} and then telescoping using the unitary relations. Doing so we get:
\begin{equation}
	\mathcal{P}_3 = \mathcal{P}_3^{(0)} + \chi \mathcal{P}_3^{(1)} + \chi {\mathcal{P}'}_3^{(1)} + \chi^2 \mathcal{P}_3^{(2)}
\end{equation}
here $\mathcal{P}_{\beta}^{(\alpha)}$ represents the term corresponding to the $\chi^{\alpha}$ factor for the expression $\mathcal{P}_{\beta}$. The term $\mathcal{P}_3^{(0)}$ immideatly simplifies to give $\frac{1}{q^2}\bra{\squarew}m\ket{\circlew}$. The next term $\mathcal{P}_3^{(1)}$ is given as:
\begin{equation}
	\mathcal{P}_3^{(1)} = \frac{\chi}{q} \,\,\, \Biggr[ \,\,\, 
	\begin{tikzpicture}[baseline=(current  bounding  box.center), scale=0.5]
		\draw[thick] (0,0) -- (3,3);
		\draw[thick] (3,0) -- (0,3);
		\draw[thick, fill = black] (3,0) circle (0.3);
		\draw[thick, fill = green, rounded corners=2pt] (1,1) rectangle (2,2);
		\draw[thick, fill = white] (3,3) circle (0.3);
		\draw[thick, fill = white] (-0.25,2.75) rectangle (0.25,3.25);
		\draw[thick, fill = white] (-0.25,-0.25) rectangle (0.25,0.25);	
	\end{tikzpicture}
	\,\,\, + \chi \,\,\,
	\begin{tikzpicture}[baseline=(current  bounding  box.center), scale=0.5]
		\draw[thick] (0,0) -- (3,3);
		\draw[thick] (3,0) -- (0,3);
		\draw[thick, fill = black] (3,0) circle (0.3);
		\draw[thick, fill = green, rounded corners=2pt] (1,1) rectangle (2,2);
		\draw[thick, fill = white] (3,3) circle (0.3);
		\draw[thick, fill = white] (-0.25,2.75) rectangle (0.25,3.25);
		\draw[thick, fill = black] (-0.25,-0.25) rectangle (0.25,0.25);	
	\end{tikzpicture}
	\,\,\, \Biggr]
	= \frac{\chi}{q}\Big( \mathcal{P}_2^{(1)} + \chi \mathcal{P}_2^{(2)} \Big)
\end{equation}
In this expression, we may now use the values obtained in Eq. \ref{P2_equation_simplified1} to get the final form of the expression. The next term ${\mathcal{P}'}_3^{(1)}$ follows the same simplification leading to the same expression. Substituting the values we are left with:
\begin{equation}
	P_3 = \frac{c}{q^3} + \frac{2\chi}{q}\Big( \frac{\sqrt{q^2 - 1}}{q} + \chi \Big( 1 - p - \frac{1}{q^2}\Big) \Big) + \chi^2 P_3^{(2)}
\end{equation}
The value of $\mathcal{P}_3^{(2)}$ has to be estimated via the Cauchy-Schwarz inequality, as performed in \cite{Dual_Dynamics_Foligno}. This is given as:
\begin{equation}
	\mathcal{P}_3^{(2)} = \,\,\,
	\begin{tikzpicture}[baseline=(current  bounding  box.center), scale=0.5]
		\foreach \i in {0,1,2}
		{
			\draw[thick] (-0.5+4*\i,-0.5) -- (5+2*\i, 5-2*\i);
			\draw[thick] (7.5-4*\i,-0.5) -- (2-2*\i, 5-2*\i);
		}
		\draw[thick, fill = black] (-0.75,-0.75) rectangle (-0.25,-0.25);
		\draw[thick, fill = green] (3.5,-0.5) circle (0.2);
		\draw[thick, fill = black] (7.5,-0.5) circle (0.2);	
		\draw[thick, fill = green, rounded corners=2pt] (1,1) rectangle (2,2);
		\draw[thick, fill = green, rounded corners=2pt] (5,1) rectangle (6,2);
		\draw[thick, fill = green, rounded corners=2pt] (3,3) rectangle (4,4);
		\draw[thick, fill = white] (5,5) circle (0.2);
		\draw[thick, fill = white] (7,3) circle (0.2);
		\draw[thick, fill = white] (9,1) circle (0.2);
		\draw[thick, fill = white] (1.75,4.75) rectangle (2.25,5.25);
		\draw[thick, fill = white] (-0.25,2.75) rectangle (0.25,3.25);
		\draw[thick, fill = white] (-2.25,0.75) rectangle (-1.75,1.25);
		
		\draw[thick, dotted] (-0.5,-2.0) -- (6.5,5.0);
		\draw[thick, dotted] (-0.5,-2.0) -- (-3.5,1.0);
		\draw[thick, dotted] (-3.5,1.0) -- (3.5,8.0);
		\draw[thick, dotted] (6.5,5.0) -- (3.5,8.0);
		
		\draw[thick, dashed] (4.0,1.5) -- (7.0,4.5);
		\draw[thick, dashed] (7.5,-2.0) -- (4.0,1.5);
		\draw[thick, dashed] (7.0,4.5) -- (10.5,1.0);
		\draw[thick, dashed] (7.5,-2.0) -- (10.5,1.0);
		
		\draw[thick] (3.5,1.0) -- (0.5,-2.0);
		\draw[thick] (3.5,1.0) -- (6.5,-2.0);
		\draw[thick] (0.5,-2.0) -- (6.5,-2.0);
	\end{tikzpicture}
	\,\,\, \le ||M||_{\infty} \sqrt{\bra{v}v\rangle} \sqrt{\bra{w}w\rangle} 
\end{equation}
Here $v$ and $w$ represent the vectors enclosed by dotted and dashed lines, whereas $M$ represents the operator enclosed by solid lines. We apply the Cauchy-Schwarz inequality, where $||M||_{\infty}$ represents the operator norm of $M$. This is equivalent to the largest singular value of $M$, or the square root of the largest eigenvalue of $M^{\dagger}M$ which is found to be $(q+c)/(q+1)$ \cite{Dual_Dynamics_Foligno}. Further, the magnitudes of vectors $v$ and $w$ are found to be dependent on the function $\eta(e_P)$ given as $\eta(e_P) = (1 - e_P)^2 + e_P^2/(q^2 - 1)$. This finally implies:
\begin{equation}
	\mathcal{P}_3^{(2)} \le \frac{q+c}{q+1}\sqrt{\eta^3(e_P)\frac{q^2 - 1}{q^2}}
	\implies \mathcal{P}_3 \le \frac{c}{q^3} + \frac{2\chi}{q}\Big( \frac{\sqrt{q^2 - 1}}{q} + \chi \Big( 1 - e_P - \frac{1}{q^2}\Big) \Big) + \frac{q+c}{q+1}\sqrt{\eta^3(e_P)\frac{q^2 - 1}{q^2}}
\end{equation}
Having performed these calculations we provide for a final example of the recursion by calculating the expression for $\mathcal{P}_4$. 
\begin{equation}
	\mathcal{P}_4 = \mathcal{P}_4^{(0)} + \chi \mathcal{P}_4^{(1)} + \chi {\mathcal{P}'}_4^{(1)} + \chi^2 \mathcal{P}_4^{(2)}
\end{equation}
\begin{equation}
	\mathcal{P}_4^{(0)} = \frac{1}{q^2}\mathcal{P}_2 = \frac{1}{q^4} + \frac{2\chi}{q^4}\sqrt{q^2 - 1} + \frac{\chi^2}{q^2}\Big(1 - e_P - \frac{1}{q^2}\Big)
\end{equation}
\begin{equation}
	\mathcal{P}_4^{(1)} = {\mathcal{P}'}_4^{(1)} = \frac{\chi}{q}\Big( \mathcal{P}_3^{(2)} + \chi \mathcal{P}_3^{(3)} \Big) \le \frac{\chi}{q} \Biggr[ \frac{\chi}{q} \Big( \frac{\sqrt{q^2 - 1}}{q} + \chi\Big( 1 - e_P - \frac{1}{q^2} \Big) \Big) + \chi \frac{q+c}{q+1}\sqrt{\eta^3(e_P)\frac{q^2 - 1}{q^2}} \Biggr]
\end{equation}
The final term in the expression, $\mathcal{P}_4^{(2)} \le \Big( \frac{q+c}{q+1}\Big)^2 \sqrt{\eta^5(e_P)\frac{q^2 - 1}{q^2}}$. Thus we find that for increasing values of $x$ we get a term which can be expressed as a power series of the parameter $\chi$. Doing so for the first few terms of $\mathcal{P}_x$ we may now have the explicit mathematical form for the lower bounds of $S_A^{(2)}(t_x)$.

\end{document}